\documentclass[prl,twocolumn,aps]{revtex4-2}

\usepackage{hyperref}
\usepackage{amsmath}
\usepackage{amsfonts}
\usepackage{amssymb}
\usepackage{epsfig}
\usepackage{graphicx}
\usepackage{braket}
\usepackage{xcolor}

\newcommand{\bea}{\begin{equation} \begin{aligned}}
\newcommand{\eea}{\end{aligned} \end{equation} }
\newcommand{\mbf}[1]{\mathbf{#1}}

\hypersetup{
    colorlinks = true,
    linkbordercolor = {white},
    citecolor=blue,
	linkcolor=blue,%
	urlcolor=black
}

\begin{document}

\title{Semi-Dirac Fermions in a Topological Metal}

\author{Yinming Shao$^{1,2}$}
\email{ys2956@columbia.edu}
\author{Seongphill Moon$^{3,4}$, A. N. Rudenko$^{5}$, Jie Wang$^{6,7,8}$, Jonah Herzog-Arbeitman$^{9}$, Mykhaylo Ozerov$^{4}$, David Graf$^{4}$, Zhiyuan Sun$^{7}$, Raquel Queiroz$^{1}$, Seng Huat Lee$^{2}$, Yanglin Zhu$^{2}$, Zhiqiang Mao$^{2}$, M. I. Katsnelson$^{5}$, B. Andrei Bernevig$^{9,10,11}$, Dmitry Smirnov$^{4}$, Andrew. J. Millis$^{1,12}$ and D. N. Basov$^{1}$}
\email{db3056@columbia.edu}

\affiliation
{$^{1}$Department of Physics, Columbia University, New York, NY, 10027, USA\\
$^{2}$Department of Physics, Pennsylvania State University, University Park, PA, 16802, USA\\
$^{3}$Department of Physics, Florida State University, Tallahassee, FL, 32306, USA\\
$^{4}$National High Magnetic Field Laboratory, Tallahassee, FL, 32310, USA\\
$^{5}$Institute for Molecules and Materials, Radboud University, Nijmegen, The Netherlands \\
$^{6}$Center of Mathematical Sciences and Applications, Harvard University, Cambridge, MA, 02138, USA \\
$^{7}$Department of Physics, Harvard University, Cambridge, MA, 02138, USA \\
$^{8}$Department of Physics, Temple University, Philadelphia, Pennsylvania, 19122, USA.\\
$^{9}$Department of Physics, Princeton University, Princeton, NJ 08544, USA\\
$^{10}$Donostia International Physics Center, P. Manuel de Lardizabal 4, 20018 Donostia-San Sebastian, Spain\\
$^{11}$IKERBASQUE, Basque Foundation for Science, Bilbao, Spain\\
$^{12}$Center for Computational Quantum Physics (CCQ), Flatiron Institute, New York, NY, 10010, USA 
}

\begin{abstract}
Topological semimetals with massless Dirac and Weyl fermions \cite{Armitage2018,Lv2021} represent the forefront of quantum materials research. In two dimensions (2D), a peculiar class of fermions that are massless in one direction and massive in the perpendicular direction was predicted sixteen years ago \cite{Dietl2008,Pardo2009,Banerjee2009}. These highly exotic quasiparticles – the semi-Dirac fermions – ignited intense theoretical and experimental interest \cite{Montambaux2009,Delplace2010,Dora2013,Huang2015,Saha2016,Roy2018,Uryszek2019,Kotov2021,Mohanta2021,yuan2016b} but remain undetected. Using magneto-optical spectroscopy, we demonstrate the defining feature of semi-Dirac fermions – \(B^{2/3}\) scaling of Landau levels – in a prototypical nodal-line metal ZrSiS \cite{Schoop2016,Hu2016}. In topological metals, including ZrSiS, nodal-lines extend the band degeneracies from isolated points to lines, loops \cite{Kim2015,Fang2015} or even chains \cite{Bzdusek2016,Yan2018,Chang2017,Chen2017} in the momentum space. With $\textit{ab initio}$ calculations and theoretical modeling, we pinpoint the observed semi-Dirac spectrum to the crossing points of nodal-lines in ZrSiS. Crossing nodal-lines exhibit a continuum absorption spectrum but with singularities that scale as \(B^{2/3}\) at the crossing. Our work sheds light on the hidden quasiparticles emerging from the intricate topology of crossing nodal-lines \cite{Bzdusek2016,Wu2019} and highlights the potential to explore quantum geometry with linear optical responses.
\end{abstract}

\maketitle

\section{I. Introduction}
Conventional 2D fermions are described by parabolic energy (E)-momentum (k) dispersion \(E(\mathbf{k})=\hbar^2 k^2/(2m)\) with effective mass \(m\). In contrast, Dirac fermions have linear dispersion \(E_D (\mathbf{k})=\hbar v_F k\) and are massless. The striking manifestations of massless Dirac fermions are revealed through the anomalous half-integer quantum Hall effect \cite{Novoselov2005,Zhang2005}, Klein tunneling \cite{Katsnelson2006,Young2009}, and \(\sqrt{B}\) scaling of Landau levels (LLs) with magnetic fields (\(B\)) \cite{Sadowski2006,Jiang2007,Orlita2008} in graphene. All these effects are observed in graphene, with the characteristic \(\sqrt{B}\) scaling provides a litmus test for Dirac quasiparticles.

\begin{figure*}[!ht]
    \centering
    \includegraphics[width=1\textwidth]{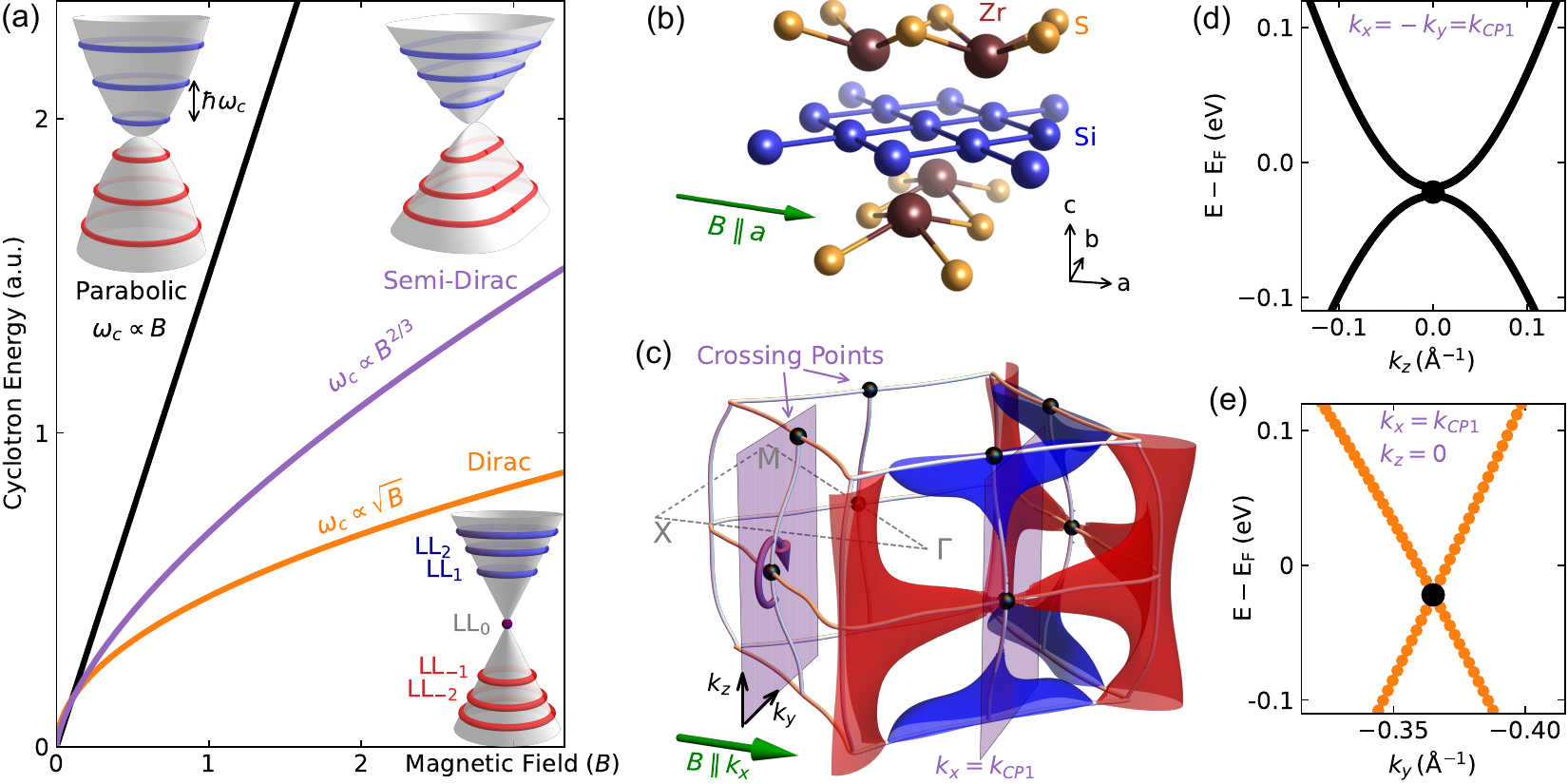}
    \caption{Semi-Dirac fermions at nodal-line crossing points in ZrSiS. (a), Cyclotron energy ($\hbar\omega_c$) as a function of magnetic field ($B$) for conventional fermions (black), Dirac fermions (orange), and semi-Dirac fermions (purple). Insets show three-dimensional plots of their band structures (grey surfaces) overlaid with Landau levels (red and blue contour lines). (b), The lattice structure of ZrSiS, showing the square lattice of Si atoms (blue) and the Zr (brown)-S (yellow) layers above and below. (c), $Ab-initio$ calculation of the Fermi surface and nodal-line structure of ZrSiS. Only the $k_x > 0$ part of the Fermi surface is shown for better visualization of the nodal-line structures (gray lines). Black spheres indicate the crossing points (CPs) of multiple nodal-lines. Purple shaded planes at $k_x = \pm k_{CP1} = \pm 0.1971 (2\pi/a)$ cross the CP1 formed by nodal-lines at $k_z=0$ and $k_x = \pm k_y$. The circular purple arrow illustrates the cyclotron motion around one of the CP1 for magnetic field (green arrow) applied along $k_x$ (a-axis of the crystal). Calculated band structure (see Methods and Supplementary Materials Sec. VII) near CP1 at $k_x=k_{CP1}$ plane shows quadratic dispersion along $k_z$ (d) and linear dispersion along $k_y$ (e), characteristic of semi-Dirac fermions.}
    \label{fig:1}
\end{figure*}

Semi-Dirac fermions, with dispersion \(E_{SD} (\mathbf{k})=\pm\sqrt{(\hbar vk_1)^2+(\hbar^2 k_2^2/(2m))^2}\) being linear in one momentum direction ($k_1$) and quadratic in the orthogonal direction ($k_2$), have been proposed to appear in materials where multiple Dirac points merge \cite{Dietl2008,Huang2015} into a semi-Dirac point. Strained graphene may be a candidate system to host semi-Dirac quasiparticles. However, the required uniaxial strain level is unrealistically large \cite{Dietl2008,Goerbig2008,Pereira2009}. Black phosphorus (BP) is proposed as another candidate semi-Dirac fermions system upon strong doping \cite{KimJ2015}. Yet the precise semi-Dirac dispersion in BP has not been established either experimentally \cite{KimJ2017} or theoretically \cite{Rudenko2015}. Thus far, the semi-Dirac dispersion \(E_{SD}\) has been experimentally explored only in synthetic platforms including honeycomb lattices of ultracold atoms \cite{Tarruell2012} and photonic resonators \cite{Bellec2013,Rechtsman2013,Real2020}. Identifying the fermionic counterpart is crucial to realize the diverse topological \cite{Huang2015,Saha2016} and correlated \cite{Roy2018,Isobe2016} phases predicted for semi-Dirac fermions, but remains challenging in 2D systems. A defining feature of semi-Dirac fermions is the unique \(B^{2/3}\) dependence \cite{Dietl2008,Banerjee2009} of inter-LL transitions (Fig.~\ref{fig:1}(a)). Here we report on the first observation of this characteristic \(B^{2/3}\) power-law in a topological metal, ZrSiS, through LL spectroscopy.

Semi-classically, a magnetic field induces cyclotron motion and the area of the cyclotron orbit at energy $E$ is $S(E) \propto E^{3/2} \sqrt{m}/v$ for the semi-Dirac dispersion $E_{SD}$~\cite{Dietl2008}. Following the Onsager quantization~\cite{Onsager1952} $S(E) = 2\pi(n+\gamma)eB/\hbar$, the characteristic $B^{2/3}$ scaling of LLs is obtained: $E_n \propto [(n+\gamma)B]^{2/3}$, where $n$ is the LL index and $\gamma$ is the phase factor $(0 \leq \gamma < 1)$. The 2D semi-Dirac spectrum can also arise as singularity points of a continuum absorption spectrum of a 3D material. For example, the LL spectrum of a nodal-ring $E_{NR} = \pm \sqrt{((k_x^2 + k_y^2 - k_0^2)/2m)^2 + v_z^2 k_z^2 }$ with field in the x-y plane exhibits a continuum absorption with a lower edge scaling as $B^{2/3}$, arising from the semi-Dirac structure (see Methods).

The prototypical nodal-line semimetal ZrSiS~\cite{Schoop2016,Hu2016,Muechler2020} (Fig.~\ref{fig:1}(b)) hosts two planar nodal-squares linked by vertical nodal-lines, forming a chain-like structure~\cite{Bzdusek2016,Yan2018} (gray lines in Fig.~\ref{fig:1}(c)) in momentum space. The low-energy physics of the ZrSiS family of nodal metals is further enriched by the Fermi energy variations along the Dirac nodal-lines~\cite{Schilling2017,Rudenko2018,Shao2020}, reflected by the coexisting electron (blue) and hole (red) pockets (Fig.~\ref{fig:1}(c)). $Ab$ $initio$ calculations and theoretical modeling show that the observed semi-Dirac fermions originate from the crossing points (CPs) of the nodal-lines in ZrSiS (black dots in Fig.~\ref{fig:1}(c)). Near the CP, the band structure at $k_x = k_{CP}$ shows quadratic (Fig.~\ref{fig:1}(d)) and linear dispersion (Fig.~\ref{fig:1}(e)) along $k_z$ and $k_y$, respectively. Under magnetic field oriented along the $a$-axis ($B \| k_x$), the cyclotron motion of electrons becomes quantized in the $(k_y,k_z)$ plane and reflects the semi-Dirac fermions through the unique LL scaling. The anticipated $B^{2/3}$ power-law in ZrSiS is robust against material complexities and can be readily identified in infrared magneto-optics experiments.

\begin{figure*}[!ht]
    \centering
    \includegraphics[width=0.922\textwidth]{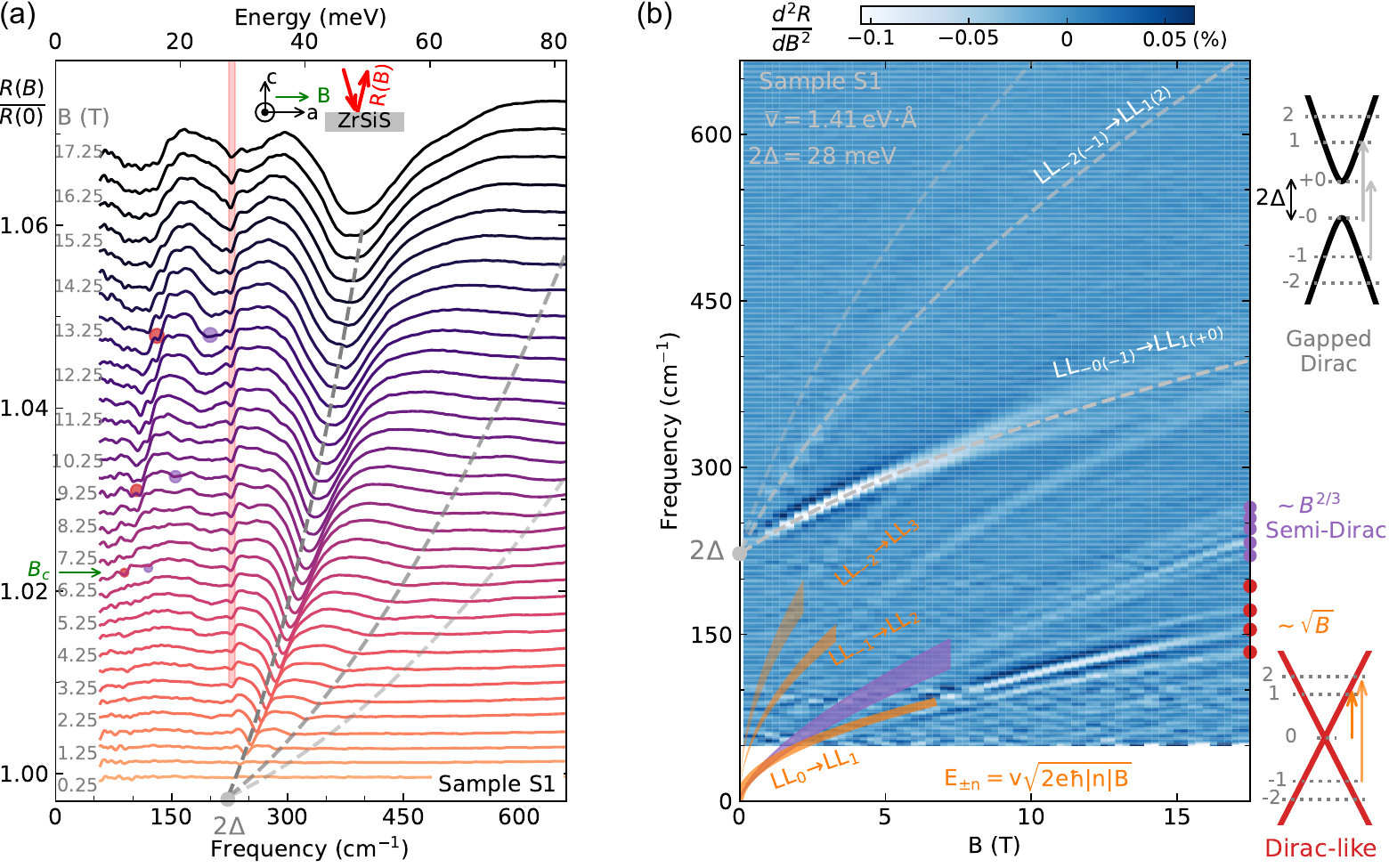}
    \caption{Landau-level spectroscopy of ZrSiS with in-plane magnetic fields. (a), Magneto-reflectance $R(\omega,B)$ normalized by zero-field reflectance $R(\omega,0 \, \text{T})$ for sample S1. Gray dashed lines mark the positions of the series of interband Landau level (LL) transitions across the spin-orbit coupling gap ($2\Delta \approx 28$ meV). The red-shaded region indicates a potential phonon feature~\cite{zhou2017,mohelsky2020,polatkan2023} that is not dispersing with increasing field. Above a critical field of $B_c \approx 7$ T, additional sub-gap transitions (red and purple dots) emerge and harden with increasing field. Inset is a schematic of the experimental configuration with near-normal incident and unpolarized light while the magnetic field is applied in-plane ($B\perp c$, $B \| a$, Voigt geometry). (b), Second derivative $d^2 R/dB^2$ data of sample S1 overlaid with model fitting (gray dashed lines) of the LL transitions across the gapped Dirac cone. The top schematic shows a gapped Dirac cone with gap $2\Delta$ and the first LL transitions $LL_{-0(-1)} \rightarrow LL_{1(+0)}$ (gray arrows). Orange lines are the guides for the sub-gap LL transitions (red dots in (a)) that follow approximately $\sqrt{B}$ scaling, characteristic of a Dirac-like fermion (bottom schematic). The sub-gap features near the purple-shaded region follow a distinct $B^{2/3}$ power-law and originate from semi-Dirac fermions.}
    \label{fig:2}
\end{figure*}

\section{II. Experimental Results}
We now proceed to the magneto-reflectance spectra $R(\omega,B)$ normalized by the zero-field data $R(\omega,0 \, \text{T})$ for ZrSiS with in-plane magnetic fields up to 17.5T (see Methods), shown in Fig.~\ref{fig:2}(a). The most prominent features are a series of dips in the reflectance spectra hardening with increasing field (gray dashed lines). For a highly metallic system like ZrSiS, the infrared reflectance approaches unity and therefore dips in $R(B)/R(0)$ correspond to absorption $A(\omega) = 1 - R(\omega)$~\cite{Schilling2017,Shao2020}. We attributed these absorption features to interband LL transitions from massive Dirac fermions ($E_{\pm n} = \sqrt{2e\hbar |n|B\bar{v}^2 + \Delta^2}$), which exhibit notable departures from the linear-in-$B$ scaling expected for fermions in parabolic bands (Fig.~\ref{fig:1}(a)). Here, $\Delta$ is half of the spin-orbit-coupling (SOC) gap~\cite{Shao2019,Shao2020,SantosCottin2021,Wyzula2022} and we find $2\Delta \approx 28$ meV, in excellent agreement with previous lower-field studies~\cite{Schilling2017,Uykur2019} and calculations~\cite{Schoop2016}. Surprisingly, above a critical field $B_c \approx 7$ T, weaker sub-gap features (red and purple dots) appear at around 100 cm$^{-1}$ and harden with increasing field. To better visualize these sub-gap structures, we report the second derivative $d^2 R/dB^2$ analysis in Fig.~\ref{fig:2}(b). The local minima of the second derivative coincide with the dips in $R(B)/R(0)$ (see Supplementary Materials Sec. III), which we identify as the LL transition energies in all analyses.

In Fig.~\ref{fig:2}(b), we show the $d^2 R/dB^2$ spectra for ZrSiS obtained with in-plane magnetic fields up to 17.5 T. The gray dashed lines denote the model calculation of interband LL transitions across the gapped Dirac cone~\cite{chen2017c,Shao2019}: $E_T = \sqrt{2e\hbar |n|B\bar{v}^2 + \Delta^2} + \sqrt{2e\hbar (|n|+1)B\bar{v}^2 + \Delta^2}$, where $n$ is the LL index and $\bar{v}$ is the averaged Fermi velocity. The intraband LL transitions~\cite{Mohelsky2023}, if present, would follow a field dependence distinct from the observed sub-gap features (see Supplementary Material Figs. S8, S25). Importantly, beyond the two series of sub-gap transitions labeled as $\sqrt{B}$ and $B^{2/3}$ (red and purple dots, respectively), two additional dispersive features are apparent above 150 cm$^{-1}$. As indicated by the thick orange lines, the dispersions of these latter features also follow approximately the $\sqrt{B}$ scaling.

\begin{figure*}[!ht]
    \includegraphics[width=1\textwidth]{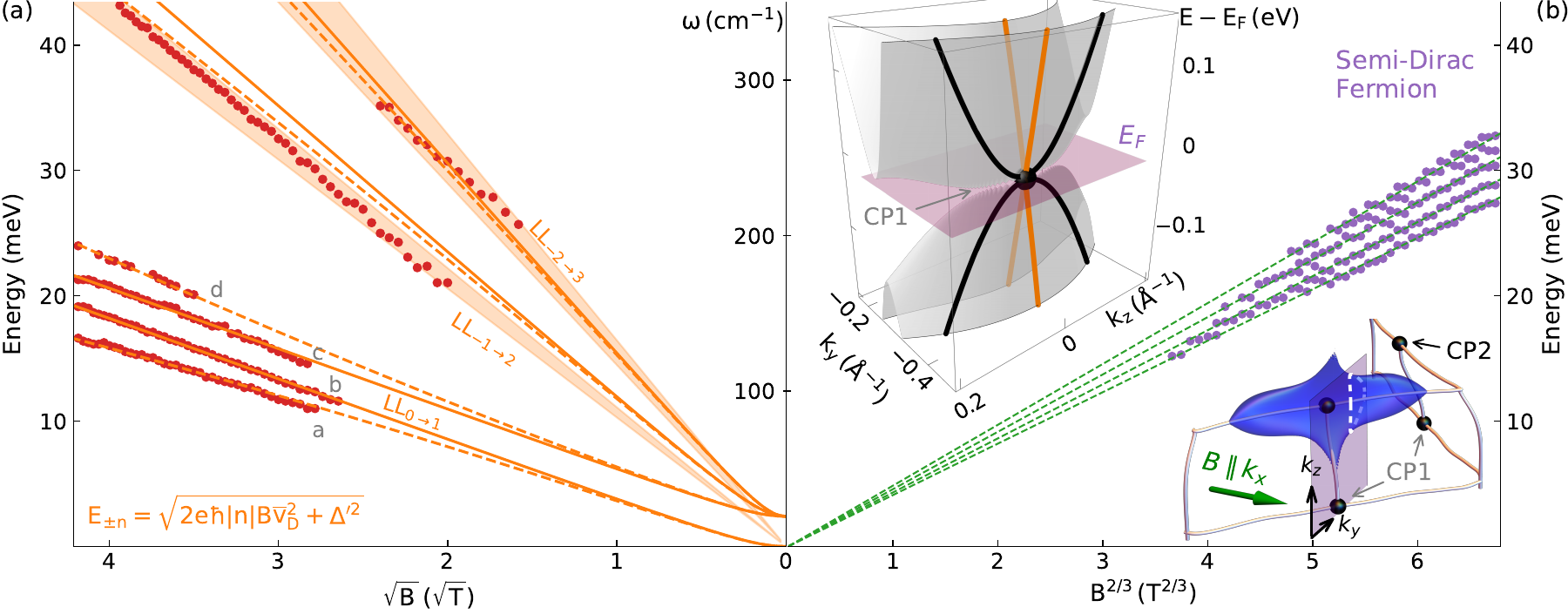}
    \caption{$\sqrt{B}$ and $B^{2/3}$ power-law behaviors of Landau levels in ZrSiS. (a) Sub-gap transition energies (red dots) in Fig.~\ref{fig:2} are plotted as a function of $\sqrt{B}$. Orange lines represent the fitting based on Dirac-like fermions with an averaged Fermi velocity $\bar{v}_D=0.88$ eVÅ, a small gap $2\Delta'=2.4$ meV, and Zeeman $g$-factor $g=2.6$. Solid and dashed lines represent the spin-conserving and spin-flip transition, respectively. Orange shaded areas indicate the uncertainties in $\bar{v}_D$ for the $LL_{-1\rightarrow2}$ ($0.91\pm0.04~\bar{v}_D$) and $LL_{-2\rightarrow3}$ ($1.02\pm0.07~\bar{v}_D$) transitions. (b), Higher-energy sub-gap transitions in Fig.~\ref{fig:2} (purple dots) are plotted as a function of $B^{2/3}$, following the exact power-law behavior expected for semi-Dirac fermions (green dashed lines). The bottom inset shows the calculated Fermi surface of ZrSiS. The shaded purple plane indicates the $k_x=k_{CP1}$ plane, which cuts through CP1 (black dot at $k_z=0$). The top inset shows the calculated band structure $E$ vs. $k_y,k_z$ for CP1 at the plane $k_x=k_{CP1}$, with linear dispersion along $k_y$ and quadratic dispersion along $k_z$, characteristic of semi-Dirac dispersion.}
    \label{fig:3}
\end{figure*}

The dipole selection rule $\delta|n|=\pm 1$ \cite{Sadowski2006,Orlita2008} for Dirac fermions dictates that the energy ratios of the lowest three interband LL transitions are: $1:1+\sqrt{2}:\sqrt{2}+\sqrt{3}$. Using a single averaged velocity of $0.82$ eVÅ, the three branches of the transitions can be approximated by the lowest three interband LL transitions from massless Dirac-like fermions (orange lines) and are labeled $LL_{0\rightarrow1}$, $LL_{-1\rightarrow2}$ and $LL_{-2\rightarrow3}$. Detailed analysis on the multiple peak splitting of $LL_{0\rightarrow1}$ below shows evidence of a small gap and spin-splitting due to Zeeman effect. On the other hand, the remaining sub-gap features (near purple shaded region in Fig.~\ref{fig:2}(b)) follow sub-linear $B$-dependence that is distinct from $\sqrt{B}$. We will confirm next that these peculiar LL transitions' field dependence scales precisely as $B^{2/3}$, a fingerprint of semi-Dirac fermions in ZrSiS.

To quantify the power-law scaling of the sub-gap features, we extract the transition energies from Fig.~\ref{fig:2}(b) (see Supplementary Material Sec. III and Figs. S7, S8) and plot them against $\sqrt{B}$ and $B^{2/3}$ in Fig.~\ref{fig:3}(a) and Fig.~\ref{fig:3}(b), respectively. Fig.~\ref{fig:3}(a) shows the experimentally determined LL transition energies (red dots) for the three groups of transitions labelled $LL_{0\rightarrow1}$, $LL_{-1\rightarrow2}$, $LL_{-2\rightarrow3}$ in Fig.~\ref{fig:2}(b). Remarkably, all these LL transitions can be understood as originating from a Dirac fermion with a small gap $2\Delta'=2.4$ meV and Zeeman-split LLs. The resulting model calculations (orange lines) show good agreement with the data (see Supplementary Material Figs.~S24,~S35). For $LL_{-2\rightarrow3}$, variation of $\bar{v}_D$ shows a logarithmic reduction with increasing $B$ field ($\bar{v}_D \propto -\ln(B)$, see Fig.~S27). Alternatively, the $LL_{-1\rightarrow2}$ and $LL_{-2\rightarrow3}$ transitions can arise from the cyclotron resonance of another gapped Dirac cone \cite{Mohelsky2023} (Fig.~S27) and the exact origin of these transitions awaits future studies.

As we alluded previously, a series of sub-gap features displays the $B^{2/3}$ scaling that is characteristic of semi-Dirac fermions (Fig.~\ref{fig:3}(b)). Fine splitting of LL transitions is also apparent and all the split peaks agree with the predicted power-law behavior (green dashed lines) for semi-Dirac fermions. These latter features are reminiscent of the spin and valley splitting of Landau levels in Dirac fermions~\cite{Chen2015,pack2020} and we discuss several possible scenarios for peak splitting in the Supplementary Materials Sec. IV and Figs. S10--S13. Due to the non-analytical nature of the LLs of semi-Dirac fermions~\cite{Dietl2008}, the selection rules have only been explored numerically for type-I semi-Dirac fermions. Nevertheless, we discuss several possibilities for the absence of additional high-order semi-Dirac LL transitions in ZrSiS in the Supplementary Materials Sec. IX and Fig. S28.

\begin{figure*}[!ht]
    \includegraphics[width=0.68\textwidth]{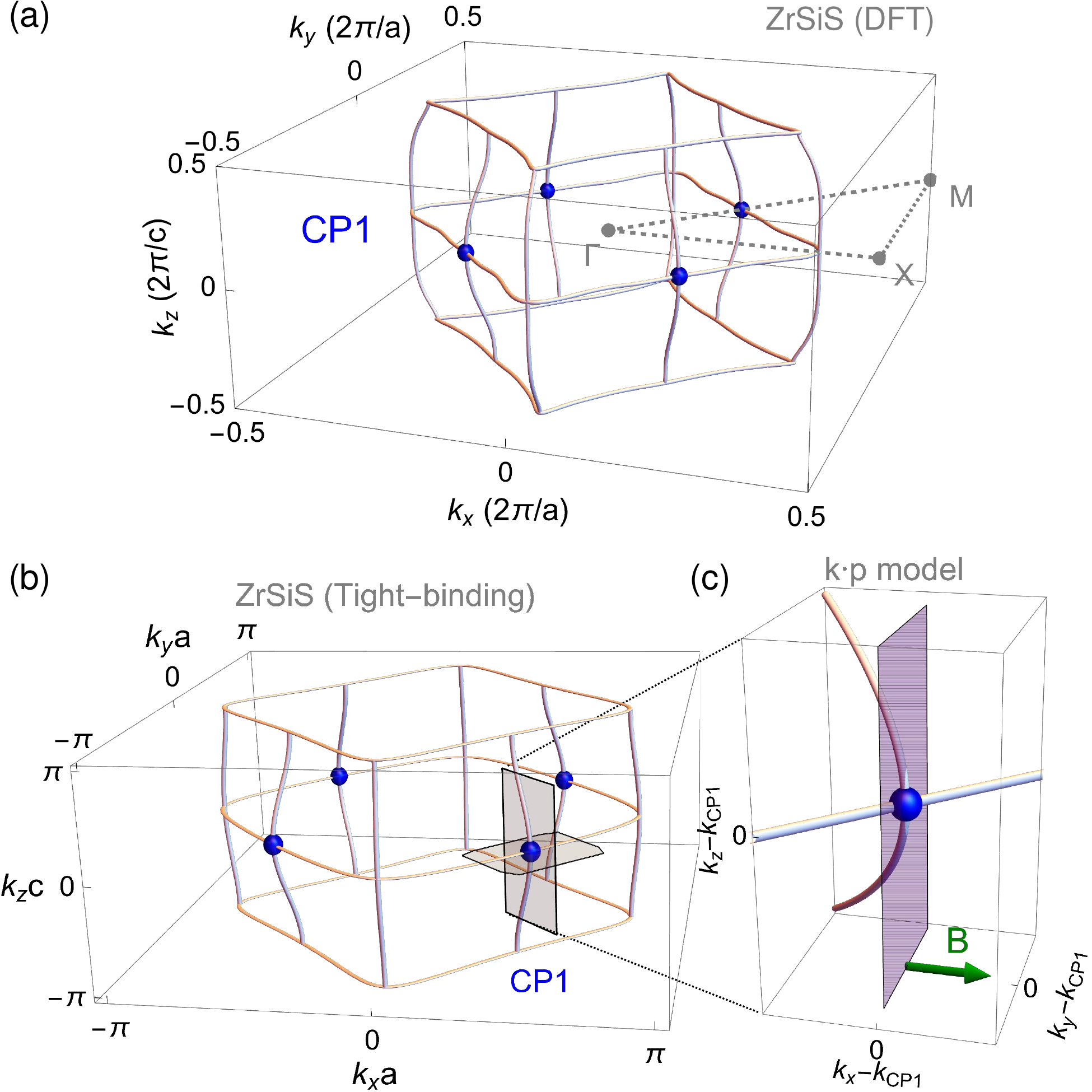}
    \caption{Nodal-line structure of ZrSiS. The 3D nodal-line structure of ZrSiS calculated using DFT (a) and the tight-binding model (b). The tight-binding model Eq.~\ref{eq:TB} (Supplementary Materials Sec. V) captures faithfully the complex structure of nodal-line crossings in DFT. Gray dashed lines connect the high symmetry points at \( k_z=0 \). Blue spheres indicate the location of the four symmetry-related crossing points at \( k_z=0 \) (CP1). The crossing of curved vertical nodal-line and straight horizontal nodal-line at CP1 hosts semi-Dirac fermions. (c), Expanding the tight-binding model (b) near CP1 and retaining the leading order terms in \( k \) leads to the two-band continuum model Eq.~\ref{eq:1}. The CP1 is formed by two nodal-lines: the vertical nodal-line defined by \( k_x=-k_y=2 \arctan \sqrt{\frac{t_{xx}}{t_{yy}}} \) with finite curvature and the horizontal nodal-line on \( k_z=0 \) with negligible curvature.}
    \label{fig:4}
\end{figure*}

\begin{figure*}[!ht]
    \includegraphics[width=0.74\textwidth]{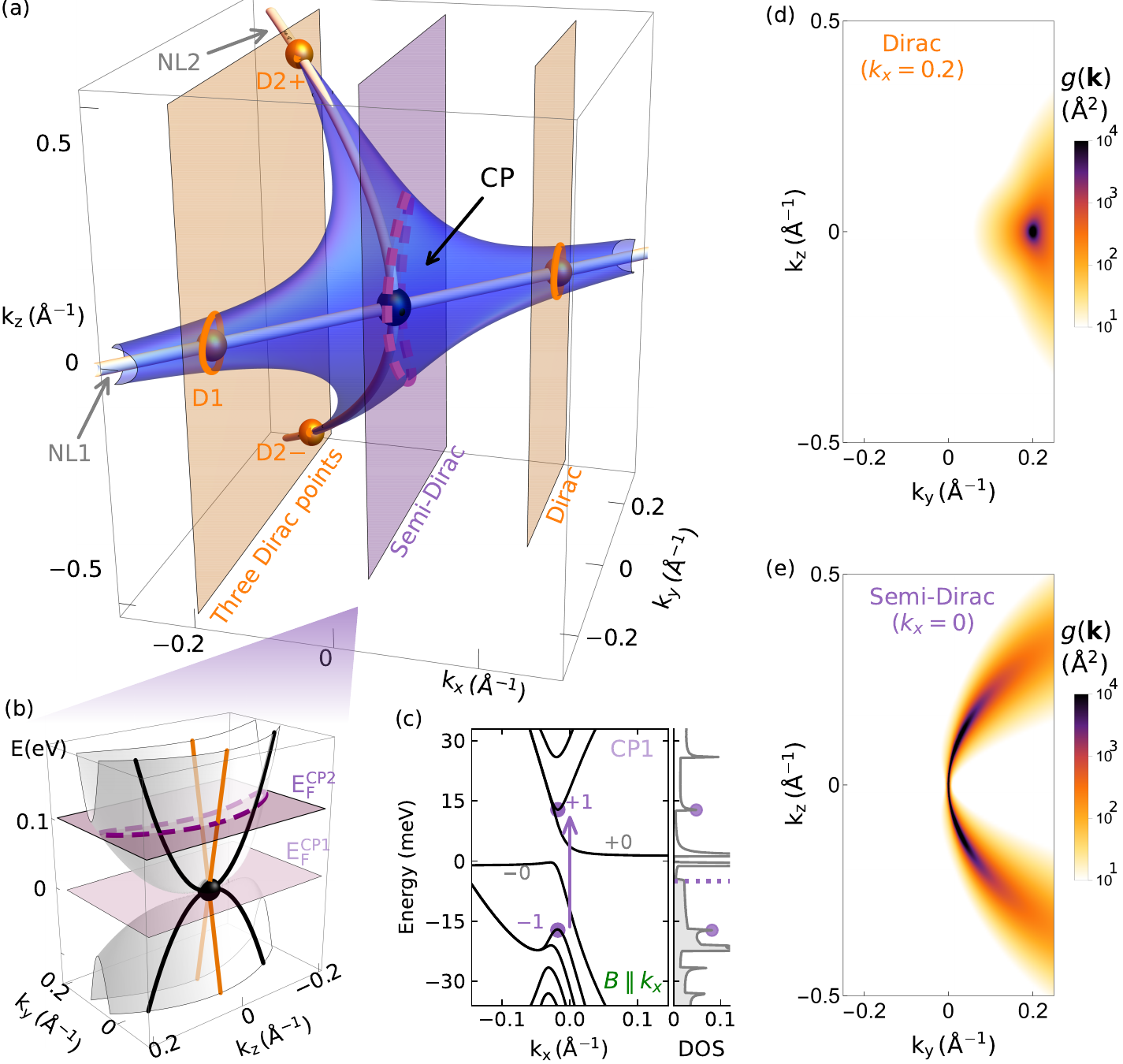}
    \caption{Semi-Dirac fermions and quantum geometry at the crossing point of two nodal lines. (a) Fermi surface (blue) of the two-band model Eq.~(\ref{eq:1}). The orange shaded plane at $k_x=-0.2\,\rm{\AA^{-1}}$ crosses the nodal line NL1 at $k_{D1}=(k_x, k_x, 0)$ and crosses NL2 at $k_{D2\pm}=(k_x, -k_x, \pm\sqrt{4mv|k_x|})$. Purple shaded plane at $k_x=0$ cuts through the CP (black sphere) of NL1 and NL2 at the origin $k=(0,0,0)$. (b), Band structure $E$ vs. $k_y,k_z$ at $k_x=0$, showing a semi-Dirac point (black sphere) as a result of the merging of three Dirac points at $k_{D1}$ and $k_{(D2\pm)}$. Purple dashed lines in (a) and (b) show the crescent-shaped Fermi surface contour of semi-Dirac fermions. Purple shaded planes represent the Fermi level for CP1 ($E_F^{CP1} \approx -5$ meV) and CP2 ($E_F^{CP2} \approx 0.1$ eV). (c), Calculated LL spectrum based on model parameters for CP1 at $B = 17.5$ T. Right panel indicates the corresponding density of states (DOS) of the LLs. Purple dashed line represent the Fermi level of CP1. Purple dots label the extremal points in the LLs and purple arrow indicate the lowest momentum-conserving transition ($LL_{-1\rightarrow+1}$), see Supplementary Materials Sec. VIII for details. (d), (e), Calculated momentum space distribution of the Fubini-Study metric $g(k)$ at a Dirac point (d, $k_x=0.2\,\rm{\AA^{-1}}$) and at the semi-Dirac point (e, $k_x=0$). Near the semi-Dirac point, $g(k)$ is nonzero and shows a stronger divergence than the Dirac case.}
    \label{fig:5}
\end{figure*}

Importantly, the semi-Dirac fermions in ZrSiS are confirmed both from \textit{ab initio} calculations and from theoretical modeling of the crossing points (CPs) of nodal-lines. As shown in the bottom inset of Fig.~\ref{fig:3}(b), there are two non-equivalent CPs in ZrSiS, labeled as CP1 (at $k_z=0$) and CP2 (at $k_z=\pi/c$). The observed $B^{2/3}$ scaling LL transition is dominated by CP1 since the energy of the CP is very close to the Fermi level, while the energy of CP2 is about 0.1 eV below the Fermi level (see Fig.~S19). The calculated semi-Dirac bands near the CP1 (top inset of Fig.~\ref{fig:3}(b)) are also asymmetric in $k_y$, in contrast to the usual type-I semi-Dirac dispersion \cite{Dietl2008}: $E_{SD}=\sqrt{k_y^4/4m^2+v^2k_z^2}$. We demonstrate below that a unique semi-Dirac fermion that originates from the merging of three Dirac points \cite{Huang2015} is realized near CP1 in ZrSiS, distinct from the merging of two Dirac points \cite{Dietl2008, Montambaux2009} realized in a single nodal-ring (Supplementary Video 1).
\begin{figure*}[!ht]
    \includegraphics[width=0.9\textwidth]{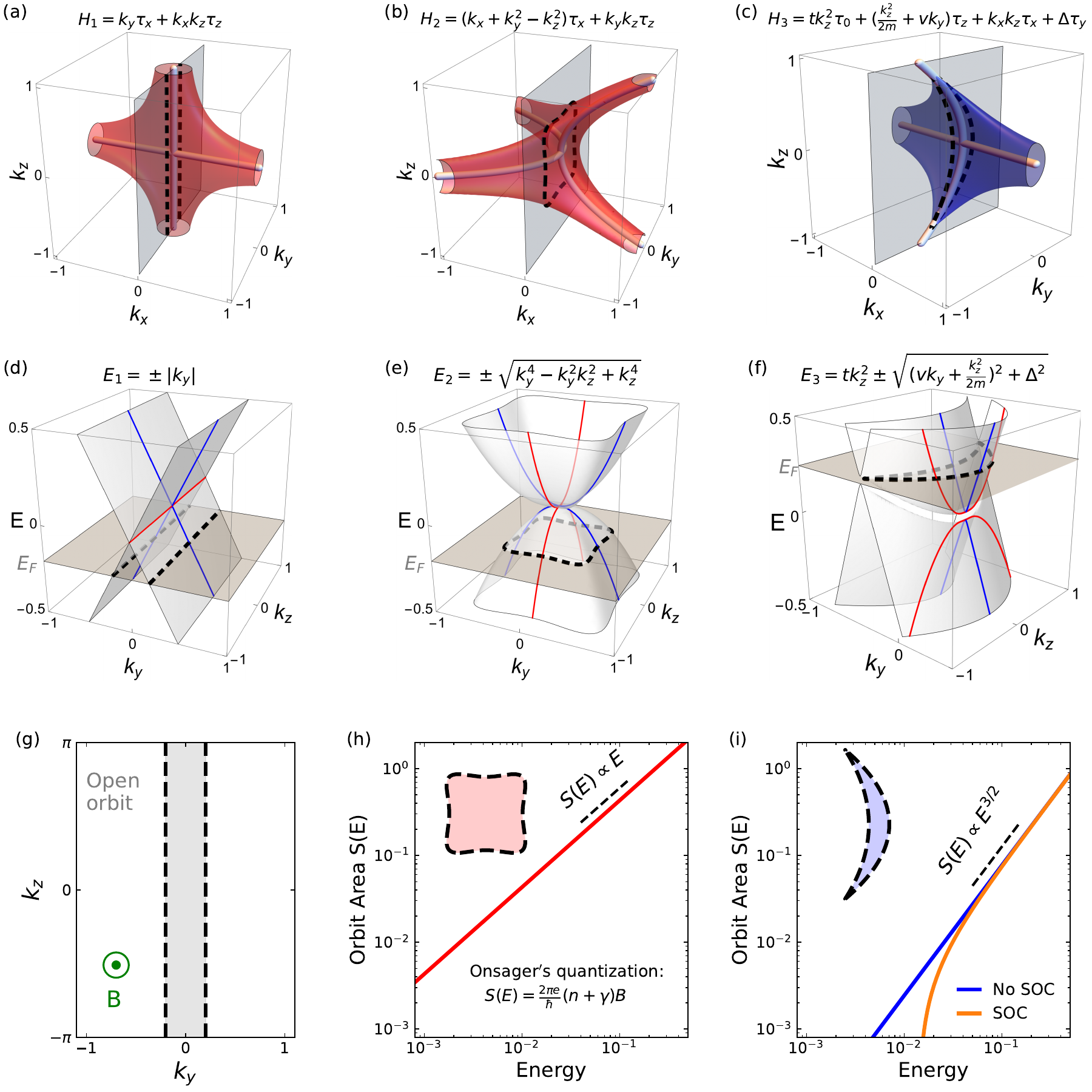}
    \caption{Nodal-line crossing point models. Three different two-band models for the crossing point of two nodal-lines at \( k=(0,0,0) \). (a), Fermi surface (FS) near the crossing of two straight nodal-lines, described by the Hamiltonian~\cite{Yan2018} \( H_1=k_y \tau_x+k_x k_z \tau_z \). (b), FS near the crossing of two parabolic nodal-lines, described by the Hamiltonian~\cite{Wu2019} \( H_2=(k_x+k_y^2-k_z^2)\tau_x+k_y k_z \tau_z \). (c), FS near the crossing of one straight and one parabolic nodal-line, described by the Hamiltonian \( H_3=tk_z^2 \tau_0+(k_z^2/2m+vk_y ) \tau_z+k_x k_z \tau_x+\Delta \tau_y \), where \( \Delta \) is half of the spin-orbit coupling (SOC) gap. The band structures at \( k_x=0 \) for the three different crossing points are shown in d, e, and f for \( H_1 \), \( H_2 \), and \( H_3 \), respectively. (g), The FS contour at \( k_x=0 \) for \( H_1 \), showing an open orbit for magnetic field \( B\parallel k_x \). (h), The area of the closed orbit \( S(E) \) at \( k_x=0 \) for \( H_2 \) increases linearly with energy. (i), \( S(E) \) increases as \( E^{3/2} \) at \( k_x=0 \) for \( H_3 \) and remains \( E^{3/2} \) for energies higher than the SOC gap. Here, \( t=0.3 \), \( v=2 \), \( m=0.5 \), \( \Delta=0.014 \). Note that $H_3$ is $e$-$h$ asymmetric and exhibits open orbits for hole-doping (Fig. S30).}
    \label{fig:6}
\end{figure*}

\section{III. Theory and Calculation}

We now turn to the theoretical interpretation of the results.  The complex 3D nodal line cage obtained in DFT reveals 8 CPs each in the $k_z = 0$ and $k_z = \pi/c$ plane. Building on the 2D Hamiltonian for square net motifs~\cite{Klemenz_2020}, we add the 3D hoppings ($t_z,t_z^\prime$) and obtain a minimal 4-band (not including spin) tight-binding model which reproduces all nodal lines and CPs, as shown in Fig.~\ref{fig:4} (see Supplementary Materials Sec. V for details):

\begin{align}
  h_{3D}(\mbf{k}) &= t_{xxz} \cos \frac{k_x a}{2} \cos \frac{k_y a}{2} (\cos k_z c- 1) \tau_1 \sigma_0  \nonumber\\
    & + t_{xyz} \sin \frac{k_x a}{2} \sin \frac{k_y a}{2} (\cos k_z c - 1) \tau_1 \sigma_1  \label{eq:TB}\\
    & + t_z (1- \cos k_z c) \sin k_x a \sin k_y a \, \tau_0 \sigma_1 \nonumber  \\ 
    &+ t_z' (\cos k_x a - \cos k_y a) \sin \! \frac{k_x a}{2} \sin \! \frac{k_y a}{2} \sin k_z c \,  \tau_2 \sigma_1 \nonumber
\end{align}

where $\tau, \sigma$ represent the sublattice and $p_x, p_y$ orbital degrees of freedom, respectively. The $a$/$c$ are the in-plane/out-of-plane lattice constants, respectively. Despite the complexity of the band structure, this model is analytically solvable, and yields closed form expressions for the nodal lines in terms of a small number of physical parameters. This model not only captures the global features of the nodal cage over the entire Brillouin zone (Fig.~\ref{fig:4}(b)), but also allows us to obtain $k.p$ models at each CP derived directly from the microscopic hoppings. In contrast to the intersection of two straight nodal-lines \cite{Yan2018} or two nodal-rings \cite{Wu2019}, the CP1 in ZrSiS is comprised of a straight nodal-line and a curved nodal-line (see Fig.~\ref{fig:1}(c) and Fig.~\ref{fig:4}), which can be described by a minimum two-band model for the CP:

\begin{equation}
    H_{CP} = t_1 k_z^2 \tau_0 + \left(\frac{k_z^2}{2m}+v k_{\perp}\right) \tau_z + t_2 k_{\parallel} k_z \tau_x + \Delta' \tau_y
    \label{eq:1}
\end{equation}

where $\tau_0$ is the identity matrix, $\tau_i  (i=x,y,z)$ are the Pauli matrices, $m$ is the effective mass along the $k_z$ direction, $v$ is the Fermi velocity along $k_{\perp}=k_x-k_y$ direction, $t_1$ controls the electron-hole asymmetry (tilt of the nodal-line along $z$), $t_2$ controls the dispersion along $k_{\parallel}=k_x+k_y$ (parallel to the in-plane nodal-line), and $\Delta'$ is half of the SOC gap. As shown in Fig.~\ref{fig:5}(a), the spectrum of Eq.~\ref{eq:1} describes the crossing of a straight in-plane nodal-line (NL1, bounded by the $k_z=0$ and $k_x=k_y$ planes) and a curved vertical nodal-line (NL2, bounded by the $k_x=-k_y$ and $k_y=k_z^2/4mv$ planes). The Fermi surface (FS) of the model (Fig.~\ref{fig:5}(a)) captures the main features of the FS in ZrSiS near the CP2 (bottom inset of Fig.~\ref{fig:3}(b)). In particular, the crescent-shape FS contour at the CP (purple dashed line) is consistent with band structure calculations of ZrSiS and further corroborated with quantum oscillation measurements (Supplementary Materials Figs. S5, S6). Under the external magnetic field along $x$, the 2D plane normal to the field at negative $k_x$ will cross NL1 once and cross NL2 twice, forming three isolated Dirac points (D1 and D2$\pm$). As the 2D plane moves to the right, the three Dirac points get closer and merge at $k_x=0$ (purple shaded plane), realizing a type-II semi-Dirac fermion \cite{Huang2015}: $H_{SD}^{\prime} = t_1 k_z^2 \tau_0 + \left(\frac{k_z^2}{2m}-v k_y\right)\tau_z + t_2 k_y k_z \tau_x$ (Fig.~\ref{fig:5}(b)).

At $k_x>0$ the spectrum becomes a single Dirac again. The unique semi-Dirac fermion described by $H_{SD}^{\prime}$ is predicted to exhibit nontrivial Berry phase and finite Chern number when a gap opens \cite{Huang2015}. This is in stark contrast with the zero Berry phase for semi-Dirac fermions formed by merging an even number of Dirac points ($\pi \times 2n$ modulo $2\pi=0$). Crossing nodal-lines in Eq.~\ref{eq:1} therefore offer a new platform for studying the rich phenomena of merging Dirac points \cite{Tarruell2012, Cao2018} (see Supplementary Video 2), where topology and correlation effects intertwine.

\begin{figure}[!ht]
    \includegraphics[width=1\columnwidth]{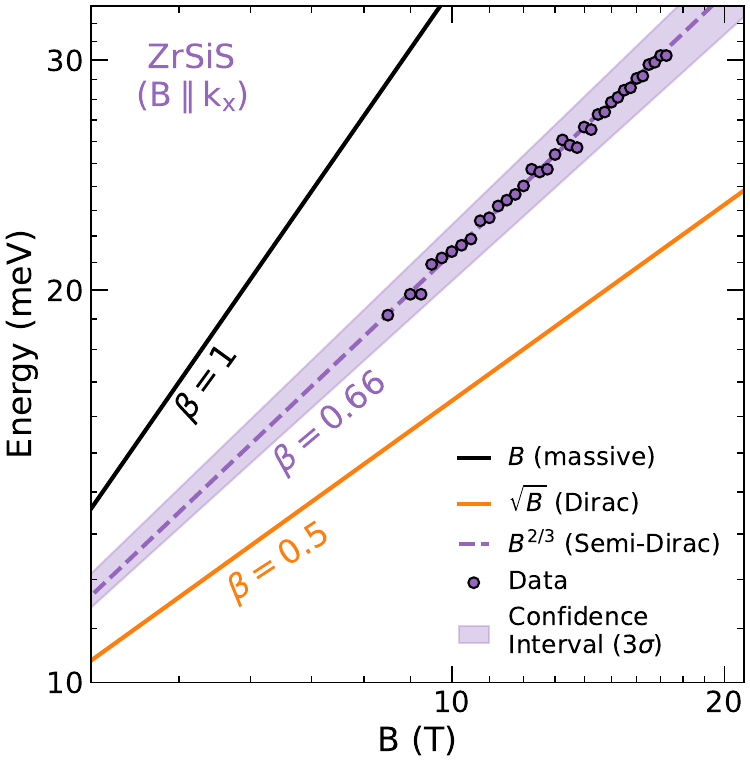}
    \caption{Comparison of the power-law of Landau level transitions for different fermions. Power-law fitting (purple dashed line) of the interband Landau level transitions (purple dots) associated with the semi-Dirac fermions in ZrSiS. Purple-shaded area indicates the \(3\sigma\) confidence interval. Orange and black lines show the power-law scaling of LLs for Dirac (\(\beta=0.5\)) and massive fermions (\(\beta=1\)), respectively.}
    \label{fig:7}
\end{figure}
We have calculated the Landau level spectrum of Eq.~\ref{eq:1} in the presence of an in-plane field directed along $x$ as a function of $k_x$, and the results are shown in Fig.~\ref{fig:5}(c). Since $k_x$ is a good quantum number for $B\|k_x$, at each $k_x$ the LL spectrum is a series of discrete levels, with level spacing determined by the projection of the constant energy contours onto the plane perpendicular to $k_x$. The LLs exhibit extremal points near $k_x=0$ and lead to peaks in the density of states (DOS), indicated by purple dots in Fig.~\ref{fig:5}(c)). Optical transitions (vertical purple arrow) between these LL singularities are observed experimentally in LL spectroscopy. We remark again that only the lowest-order momentum-conserving LL transitions ($LL_{-1\rightarrow+1}$) have been observed in this work (Supplementary Materials Sec. IX). Higher-order LL transitions are in general weaker and may be forbidden by selection rules (see e.g. ~\cite{zhou2021} for type-I semi-Dirac fermions). We further demonstrate the $B^{2/3}$ scaling of the LLs at the CP using both semi-classical quantization below and full LL calculation with SOC (Supplementary Materials Sec. VIII). Apart from enhanced DOS, we remark that the semi-Dirac fermions realized by Eq.~\ref{eq:1} also show stronger divergence in the quantum metric $g(k)$ compared to Dirac fermions (see Fig.~\ref{fig:5}(d)). In Fig.~\ref{fig:5}(d), we compared the calculated $k$-space distribution of $g(k)$ at $k_x=0$ (semi-Dirac) and $k_x=0.2\,\rm{\AA^{-1}}$ (Dirac) based on Eq.~\ref{eq:1} (Supplementary Materials Sec. VI). We remark that the effects of quantum geometry in solids have been discussed mostly theoretically in terms of nonlinear optical response \cite{Ahn2020, Ahn2022}, but recent~\cite{onishi2024a} and classic work \cite{Souza2000, Resta2011} has connected the linear response to the integrated Fubini-Study metric. Our experimental observation of semi-Dirac fermions highlights the potential to explore quantum geometry from novel quasiparticles using linear optical and magneto-optical response.

Before concluding, we comment on the robustness of the observed $B^{2/3}$ scaling of Landau levels from crossing nodal-lines in ZrSiS. There are several different ways that two nodal-lines cross in momentum space~\cite{Yan2018,Wu2019}. We illustrate three of the lowest orders crossing in Fig.~\ref{fig:6}, where the two nodal-lines are (a) both straight~\cite{Yan2018} ($H_1=k_y \tau_x+k_x k_z \tau_z$), (b) both parabolic~\cite{Wu2019} \( H_2=(k_x+k_y^2-k_z^2)\tau_x+k_y k_z \tau_z \), or (c) consisting of one straight and one parabolic nodal-line \( H_3=tk_z^2 \tau_0+(k_z^2/2m+vk_y ) \tau_z+k_x k_z \tau_x+\Delta \tau_y \), where \( \Delta \) is half of the SOC gap. Under an external magnetic field along $k_x$, the cyclotron motions of electrons will be quantized in the 2D plane normal to the magnetic field direction (constant $k_x$ planes). The band-structures corresponds to the three models in Figs.~\ref{fig:6}(a)-(c) at the plane $k_x=0$ are shown in Figs.~\ref{fig:6}(d)-(f), respectively. The cyclotron motions for charge carriers from the three different nodal-line crossing points can now be assessed by their distinct Fermi surface cross-sections at the Fermi level (black dashed lines in Fig.~\ref{fig:6}(d)-(f)). Importantly, the bandstructure at the CP from $H_3$ (Fig.~\ref{fig:6}(f)) is electron-hole asymmetric (see also Fig. S30), disctinct from $H_1$ and $H_2$.

We can evaluate the Landau Level scaling from a semi-classical approach for these three different crossing points. First, it is evident that the crossing of two straight nodal-lines ($H_1$) will leads to an open orbit at the CP (Fig.~\ref{fig:6}(g)), and therefore will not give rise to quantized energy levels. For the crossing of two-parabolic nodal-lines ($H_2$), the area of the cyclotron orbit $S(E)$ increase linearly with energy $E$ (Fig.~\ref{fig:6}(h)). According to Onsager quantization relation~\cite{Onsager1952}: $S(E) = 2\pi(n+\gamma)eB/\hbar$ where $n$ is the Landau level index and $\gamma$ is the phase factor, the LL scaling will be linear in $B$. In contrast, as derived in the Methods section, the cyclotron orbit area increase as $E^{3/2}$ (Fig.~\ref{fig:6}(i)) for the CP of a straight and parabolic nodal-line ($H_3$), which results in a LL scaling that is strictly $B^{2/3}$. This semi-classical analysis is consistent with full quantum mechanical calculations the CP of a straight and parabolic nodal-lines, which also exhibits the $B^{2/3}$ LL scaling (see Supplementary Material Sec. VIII and Fig. S23).

The generic nature of our nodal-line crossing point model and analysis (Figs.~\ref{fig:4}-~\ref{fig:6}) guarantees the robustness of the experimental observable even in real and complex materials such as ZrSiS. In Fig.~\ref{fig:7}, we show the power-law fitting of one of the Landau level transitions from semi-Dirac fermions (purple dots) with function $A \cdot B^\beta$ and the corresponding $3\sigma$ confidence interval (purple shaded region). The data points are consistent with the $B^{2/3}$ scaling and statistically different from $B$ (conventional massive fermions), $B^{4/5}$ (three-quarter Dirac fermions~\cite{Kishigi2017}) and $B^{1/2}$ (Dirac fermions).

Using Landau-level spectroscopy, we uncovered the semi-Dirac fermions inside bulk ZrSiS under in-plane magnetic fields. Our results demonstrate novel magnetic field effect in ZrSiS not explored in previous experiments utilizing out-of-plane magnetic fields \cite{Pezzini2018, Muller2020, Gudac2022}. In contrast to the conventional expectation of 2D electrons at the surface/interface of a layered material, the observed semi-Dirac fermions reside within planes perpendicular to the atomic layers of ZrSiS and originate from the vicinity of points where nodal-lines cross. The crossing point of nodal-lines in ZrSiS offers a unique and generic platform for realizing semi-Dirac fermions through the merging of three Dirac points. Our findings advance the understanding of exotic 2D electrons in natural bulk crystals, establish the existence of novel quasiparticles associated with crossing nodal lines in momentum space \cite{Bzdusek2016, Wu2019}, and open new directions in exploring quantum geometry and topological effects in metals.

\section{Acknowledgement}
\begin{acknowledgments}
Magneto-optical spectroscopy of ZrSiS is supported by NSF-DMR 2210186. Research on the electrodynamics of semi-Dirac quasiparticles, including the theoretical analysis of Landau Levels spectra at Columbia, is supported as part of Programmable Quantum Materials, an Energy Frontier Research Center funded by the U.S. Department of Energy (DOE), Office of Science, Basic Energy Sciences (BES), under award DE-SC0019443. D.N.B. is Moore Investigator in Quantum Materials EPIQS GBMF9455. Support for crystal growth and characterization at Penn State was provided by the National Science Foundation through the Penn State 2D Crystal Consortium-Materials Innovation Platform (2DCC-MIP) under NSF Cooperative Agreement DMR-1539916 and NSF-DMR 2039351. The National High Magnetic Field Laboratory is supported by the National Science Foundation through NSF/DMR-1644779, NSF/DMR-2128556 and the State of Florida. The work of A.N.R. and M.I.K. was supported by the European Union's Horizon 2020 research and innovation program under European Research Council Synergy Grant 854843 ``FASTCORR". The work of M.I.K. was further supported by the Dutch Research Council (NWO) via the ``TOPCORE" consortium. The Flatiron Institute is a division of the Simons Foundation.	
\end{acknowledgments}

\section{Methods}
\subsection{Voigt magneto-optical spectroscopy and the absence of surface states.}
High-field magneto-optical measurements were performed at $T \approx 5$ K under Voigt geometry ($B \perp c$ and B $\parallel a$) at the National High Magnetic Field Laboratory. A Bruker Vertex 80V FTIR spectrometer combined with a $17.5$ T superconducting magnet was utilized to record the reflectance spectra of ZrSiS at zero and high magnetic field. Infrared beam from a Globar lamp was focused on the (001) surface of ZrSiS crystal. The typical size of ZrSiS crystals used in our measurements is $\sim 4$ mm $\times$ $4$ mm $\times$ $0.3$ mm and the spot size of the infrared beam is smaller than the lateral size of the sample. 

We remark that there are various surface states observed by angular-resolved photoemission measurements \cite{Schoop2016, Muechler2020, Topp2017, Nakamura2019} from the (001) surface of the ZrSiS family of topological semimetals. These surface states are strictly confined to the top and bottom surfaces of ZrSiS and do not contribute to the out-of-plane orbital motion when the magnetic field is applied in-plane. Therefore, we conclude that the previously reported surface states from (001) surface of ZrSiS cannot explain the nearly massless Dirac fermions observed in our Voigt geoemtry magneto-optical data.

\subsection{Density functional theory calculation} DFT calculations were carried out using the plane-wave pseudopotential method as implemented in the Quantum ESPRESSO simulation package \cite{Giannozzi2009, Giannozzi2017}. Norm-conserving pseudopotentials \cite{Prandini2018} were used in conjunction with the local density approximation for exchange-correlation potential. An energy cutoff of $70$ Ry for the plane-waves and a convergence threshold of $10^{-12}$ Ry were used for the self-consistent solution of the Kohn-Sham equations. The Brillouin zone was sampled by a $14 \times 14 \times 6$ Monkhorst-Pack \cite{Monkhorst1976} k-point mesh. Lattice constants were relaxed, resulting in $a = 3.4665$ Å and $c = 7.9148$ Å. The atomic structure within the unit cell was relaxed until the residual forces were less than $10^{-4}$ Ry/bohr.

To ensure the numerical accuracy of the Fermi surface calculations and related properties, an interpolation scheme based on the maximally localized Wannier functions (MLWF) \cite{Marzari2012} was used. For this purpose, we used the wannier90 code \cite{Mostofi2014} to construct an extended tight-binding Hamiltonian for ZrSiS in the MLWF basis, which included the $3s$ and $3p$ states for Si and S as well as the $4s$ and $4d$ states for Zr. The interpolated band energies ensure a correct description of the DFT band structure within the range of at least $\pm 10$ eV relative to the Fermi energy. The Wannier-interpolated band energies were calculated on a $(300 \times 300 \times 300)$ k-point mesh, and were further used to calculate the de Haas-van Alphen frequencies using the Supercell k-space Extremal Area Finder (SKEAF) code \cite{Rourke2012}.

\subsection{Quantum oscillations in ZrSiS with torque magnetometry measurements} The torque magnetometry measurements up to $14$ T were performed using a piezoresistive cantilever in a superconducting magnet equipped with a variable temperature insert (VTI). A single crystal of ZrSiS is fixed to the end of a $0.30$ mm cantilever arm with vacuum grease. A jet of Helium-4 gas from the inlet of the VTI onto the cryostat kept the samples at a constant temperature of $1.6$ K during the measurements. There are two resistive elements on the cantilever, one of which is located at the base of the arm and experiences strain with a change in the sample magnetization. The second resistive element is not affected by the torque but mimics the temperature and magnetic field dependence of the first. These are combined with two more resistors at room temperature to form a Wheatstone bridge that can be balanced at low temperatures before changing the magnetic field. A small current is applied across the bridge circuit and the measured voltage records the changing torque $\tau$ created by the de Haas-van Alphen (dHvA) effect.

\subsection{Semi-Dirac fermions in the nodal-ring model}
To illustrate the semi-Dirac fermions in the nodal-ring model, we consider the nodal-ring Hamiltonian \cite{Yang2018,Oroszlany2018}: 
\begin{equation}
    H_R (\mathbf{k}) = (k_x^2 + k_y^2 - k_0^2)\tau_x + v k_z \tau_z, \label{eq:2}
\end{equation}
 
where $v$ is the Fermi velocity along $z$, $\tau_0$ is the identity matrix, and $\tau_x, \tau_z$ are the Pauli matrices for orbitals. The corresponding energy spectrum is:
\begin{equation}
    E_R (\mathbf{k}) = \pm\sqrt{(k_x^2 + k_y^2 - k_0^2)^2 + v^2 k_z^2} \label{eq:3}
\end{equation}
and a Dirac nodal-ring at $k_z=0$ with radius $k_0 = \sqrt{k_x^2 + k_y^2}$ can be identified. This nodal-ring marks a protected crossing of two bands along a ring in momentum space and at any point on the ring, the spectrum is Dirac. Remarkably, at $k_x = \pm k_0$, the projection of constant energy contours onto the $k_y-k_z$ plane yields contours of the semi-Dirac form \cite{Montambaux2009,Oroszlany2018} $E_{SD} = \pm\sqrt{k_y^4/4m^2 + v^2 k_z^2}$ that describes a massive fermion along $k_y$ ($m=0.5$) and massless Dirac fermion along $k_z$ (see Supplementary Video 1). Under the in-plane magnetic field, the area of the cyclotron orbit at energy $E$ is $S(E) \propto E^{3/2} \sqrt{m/v}$, which leads to the $B^{2/3}$ scaling of LLs \cite{Dietl2008, Oroszlany2018}. Here, we briefly describe the solution of the Hamiltonian in an in-plane magnetic field $B$ directed along $x$ that reveals the main features of the absorption spectrum. Choosing the Landau gauge $\mathbf{A} = (0, Bz, 0)$ and rescaling the coordinate $z \rightarrow l_B u$ with magnetic length $l_B = k_0/\sqrt{B}$ and defining field scale $B_0 = k_0^2$ and rescaled field $b = B/B_0$, we have $H_R (\mathbf{k}) = 1/l_B (V(u)\tau_x - i\partial_u \tau_z)$ with $V(u) = 1/\sqrt{b}(-(1 - (k_x^2)/(k_0^2)) + bu^2)$. From this form of $V$ one can see that for $(1 - (k_x^2)/(k_0^2)) \neq 0$, $V$ is minimized at a particular $u^2$ expanding around the two $u$ values that give Dirac spectra, while when $k_x^2 = k_0^2$ we have a semi-Dirac equation. We have solved the corresponding Schrödinger equation numerically by observing that at each $k_x$, $H_R^2$ can be diagonalized trivially yielding two second-order differential equations that we solve by discretization. The current operator can then be obtained as a differential operator and applied to the solutions. Integration of the resulting absorption spectrum over $k_x$ gives the continuum absorption, with singularities at the upper edge scaling as $B^{1/2}$ and lower edge as $B^{2/3}$, consistent with previous results \cite{Oroszlany2018}. While the nodal-ring model is not directly relevant to the interpretation of our experimental results, it does display the generic features of an absorption continuum with upper and lower edges scaling differently with $B$, and with the characteristic Dirac and semi-Dirac behaviors.

\subsection{Semiclassical quantization of the crossing nodal-line model}

To obtain the Landau level scaling of the semi-Dirac fermion at the CP of nodal-lines, we consider the following Hamiltonian:
\begin{equation}
    H_3 = tk_z^2 \tau_0 + \left(\frac{k_z^2}{2m} - vk_y\right) \tau_z + k_x k_z \tau_x,\label{eq:CP}
\end{equation}

which has an extremal Fermi surface at $k_x=0$ for $B\parallel k_x$. This model also describes the crossing of one straight and one curved nodal-line and is related to the model Eq.~\ref{eq:1} by a 45-degree in-plane rotation. The corresponding eigenvalue is given by:
\begin{equation}
    E_3 = tk_z^2 \pm \sqrt{\frac{k_z^4}{4m^2} - \frac{vk_z^2 k_y}{m} + v^2 k_y^2 + k_x^2 k_z^2}
\end{equation}

For a magnetic field $B$ applied along $k_x$, the eigenvalue spectrum has extremal values at $k_x=0$. Without loss of generality, we consider the Fermi surface of the electron pockets ($m>0$) and choose $v>0$, $E>0$, and $0<t<(2m)^{-1}$. The cross-section of the electron Fermi surface (FS) at constant energy $E$ is then determined by:
\begin{equation}
    E = tk_z^2 \pm \frac{k_z^2}{2m} - vk_y, 
\end{equation}

which is bounded by two parabolic curves:
\begin{equation}
    k_y^+ = \frac{E}{v} - \frac{t - \frac{1}{2m}}{v} k_z^2 \quad \text{and} \quad k_y^- = -\frac{E}{v} + \frac{t + \frac{1}{2m}}{v} k_z^2.
\end{equation}

These two parabolic curves intersect at $k_z^2 = E/t \equiv a^2$, forming a crescent-shaped closed contour for electron orbits (see inset of Fig.~\ref{fig:6}(i)). If $k_z^2 > E/t$, the hole FS is obtained with $E = tk_z^2 - \frac{k_z^2}{2m} - vk_y$. The area of the electron FS contour can be obtained by evaluating the difference between the areas of the parabolic segments of $k_y^-$ and $k_y^+$:

\begin{align}
S(E) &= S^-(E) - S^+(E) = \int_{-a}^{a} (k_y^+ - k_y^-) dk_z \\ \nonumber
&= \int_{-a}^{a} 2E/v \left(1 - \frac{k_z^2}{a^2}\right) dk_z = \frac{8}{3} \frac{E^{3/2}}{\sqrt{t} v}. 
\label{eq:onsager}
\end{align}

Following Onsager’s quantization relation $S(E) = 2\pi(n+\gamma)eB/\hbar$ where $n$ is the Landau level index and $\gamma$ is the phase factor, we obtain the $B^{2/3}$ scaling of LLs: $E_n \propto (n+\gamma)^{2/3} B^{2/3}$.

In the presence of finite SOC, the degeneracy at the semi-Dirac point is lifted but the characteristic band dispersions and LL scaling recover at energies higher than the SOC gap. We illustrate the case with SOC using the following Hamiltonian based on Eq.~\ref{eq:CP}:
\begin{equation}
    H_3' = tk_z^2 \tau_0 + \left(\frac{k_z^2}{2m} - vk_y\right) \tau_z + k_x k_z \tau_x + \Delta\tau_y,
\end{equation}

where $2\Delta$ is the size of the SOC gap. The area of the electron FS contour with SOC becomes:

\begin{multline}
S_{SOC}(E) = \frac{8}{3} \frac{\sqrt{E+\Delta}}{\sqrt{t} v} \\
\times \left( e_E\left[1 - 2\frac{\Delta}{E+\Delta}\right]E - e_K\left[1 - 2\frac{\Delta}{E+\Delta}\right]\Delta \right), \label{area_soc}
\end{multline}

where $e_E[x]$ and $e_K[x]$ denote the complete elliptic integral of the second kind and first kind, respectively. Eq.~\ref{area_soc} can be solved numerically and the resulting $S_{SOC}(E)$ is shown in Fig.~\ref{fig:6}(i), where the $S_{SO}(E) \propto E^{3/2}$ behavior is recovered at energies higher than around $2\Delta$.


\begin{thebibliography}{80}%
\makeatletter
\providecommand \@ifxundefined [1]{%
 \@ifx{#1\undefined}
}%
\providecommand \@ifnum [1]{%
 \ifnum #1\expandafter \@firstoftwo
 \else \expandafter \@secondoftwo
 \fi
}%
\providecommand \@ifx [1]{%
 \ifx #1\expandafter \@firstoftwo
 \else \expandafter \@secondoftwo
 \fi
}%
\providecommand \natexlab [1]{#1}%
\providecommand \enquote  [1]{``#1''}%
\providecommand \bibnamefont  [1]{#1}%
\providecommand \bibfnamefont [1]{#1}%
\providecommand \citenamefont [1]{#1}%
\providecommand \href@noop [0]{\@secondoftwo}%
\providecommand \href [0]{\begingroup \@sanitize@url \@href}%
\providecommand \@href[1]{\@@startlink{#1}\@@href}%
\providecommand \@@href[1]{\endgroup#1\@@endlink}%
\providecommand \@sanitize@url [0]{\catcode `\\12\catcode `\$12\catcode
  `\&12\catcode `\#12\catcode `\^12\catcode `\_12\catcode `\%12\relax}%
\providecommand \@@startlink[1]{}%
\providecommand \@@endlink[0]{}%
\providecommand \url  [0]{\begingroup\@sanitize@url \@url }%
\providecommand \@url [1]{\endgroup\@href {#1}{\urlprefix }}%
\providecommand \urlprefix  [0]{URL }%
\providecommand \Eprint [0]{\href }%
\providecommand \doibase [0]{https://doi.org/}%
\providecommand \selectlanguage [0]{\@gobble}%
\providecommand \bibinfo  [0]{\@secondoftwo}%
\providecommand \bibfield  [0]{\@secondoftwo}%
\providecommand \translation [1]{[#1]}%
\providecommand \BibitemOpen [0]{}%
\providecommand \bibitemStop [0]{}%
\providecommand \bibitemNoStop [0]{.\EOS\space}%
\providecommand \EOS [0]{\spacefactor3000\relax}%
\providecommand \BibitemShut  [1]{\csname bibitem#1\endcsname}%
\let\auto@bib@innerbib\@empty
\bibitem [{\citenamefont {Armitage}\ \emph {et~al.}(2018)\citenamefont
  {Armitage}, \citenamefont {Mele},\ and\ \citenamefont
  {Vishwanath}}]{Armitage2018}%
  \BibitemOpen
  \bibfield  {author} {\bibinfo {author} {\bibfnamefont {N.~P.}\ \bibnamefont
  {Armitage}}, \bibinfo {author} {\bibfnamefont {E.~J.}\ \bibnamefont {Mele}},\
  and\ \bibinfo {author} {\bibfnamefont {A.}~\bibnamefont {Vishwanath}},\
  }\bibfield  {title} {\bibinfo {title} {Weyl and dirac semimetals in
  three-dimensional solids},\ }\href
  {https://doi.org/10.1103/RevModPhys.90.015001} {\bibfield  {journal}
  {\bibinfo  {journal} {Rev. Mod. Phys.}\ }\textbf {\bibinfo {volume} {90}},\
  \bibinfo {pages} {015001} (\bibinfo {year} {2018})}\BibitemShut {NoStop}%
\bibitem [{\citenamefont {Lv}\ \emph {et~al.}(2021)\citenamefont {Lv},
  \citenamefont {Qian},\ and\ \citenamefont {Ding}}]{Lv2021}%
  \BibitemOpen
  \bibfield  {author} {\bibinfo {author} {\bibfnamefont {B.~Q.}\ \bibnamefont
  {Lv}}, \bibinfo {author} {\bibfnamefont {T.}~\bibnamefont {Qian}},\ and\
  \bibinfo {author} {\bibfnamefont {H.}~\bibnamefont {Ding}},\ }\bibfield
  {title} {\bibinfo {title} {Experimental perspective on three-dimensional
  topological semimetals},\ }\href
  {https://doi.org/10.1103/RevModPhys.93.025002} {\bibfield  {journal}
  {\bibinfo  {journal} {Rev. Mod. Phys.}\ }\textbf {\bibinfo {volume} {93}},\
  \bibinfo {pages} {025002} (\bibinfo {year} {2021})}\BibitemShut {NoStop}%
\bibitem [{\citenamefont {Dietl}\ \emph {et~al.}(2008)\citenamefont {Dietl},
  \citenamefont {Piéchon},\ and\ \citenamefont {Montambaux}}]{Dietl2008}%
  \BibitemOpen
  \bibfield  {author} {\bibinfo {author} {\bibfnamefont {P.}~\bibnamefont
  {Dietl}}, \bibinfo {author} {\bibfnamefont {F.}~\bibnamefont {Piéchon}},\
  and\ \bibinfo {author} {\bibfnamefont {G.}~\bibnamefont {Montambaux}},\
  }\bibfield  {title} {\bibinfo {title} {New magnetic field dependence of
  landau levels in a graphenelike structure},\ }\href
  {https://doi.org/10.1103/PhysRevLett.100.236405} {\bibfield  {journal}
  {\bibinfo  {journal} {Phys. Rev. Lett.}\ }\textbf {\bibinfo {volume} {100}},\
  \bibinfo {pages} {236405} (\bibinfo {year} {2008})}\BibitemShut {NoStop}%
\bibitem [{\citenamefont {Pardo}\ and\ \citenamefont
  {Pickett}(2009)}]{Pardo2009}%
  \BibitemOpen
  \bibfield  {author} {\bibinfo {author} {\bibfnamefont {V.}~\bibnamefont
  {Pardo}}\ and\ \bibinfo {author} {\bibfnamefont {W.~E.}\ \bibnamefont
  {Pickett}},\ }\bibfield  {title} {\bibinfo {title} {Half-metallic
  semi-dirac-point generated by quantum confinement in tio2/vo2
  nanostructures},\ }\href {https://doi.org/10.1103/PhysRevLett.102.166803}
  {\bibfield  {journal} {\bibinfo  {journal} {Phys. Rev. Lett.}\ }\textbf
  {\bibinfo {volume} {102}},\ \bibinfo {pages} {166803} (\bibinfo {year}
  {2009})}\BibitemShut {NoStop}%
\bibitem [{\citenamefont {Banerjee}\ \emph {et~al.}(2009)\citenamefont
  {Banerjee}, \citenamefont {Singh}, \citenamefont {Pardo},\ and\ \citenamefont
  {Pickett}}]{Banerjee2009}%
  \BibitemOpen
  \bibfield  {author} {\bibinfo {author} {\bibfnamefont {S.}~\bibnamefont
  {Banerjee}}, \bibinfo {author} {\bibfnamefont {R.~R.~P.}\ \bibnamefont
  {Singh}}, \bibinfo {author} {\bibfnamefont {V.}~\bibnamefont {Pardo}},\ and\
  \bibinfo {author} {\bibfnamefont {W.~E.}\ \bibnamefont {Pickett}},\
  }\bibfield  {title} {\bibinfo {title} {Tight-binding modeling and low-energy
  behavior of the semi-dirac point},\ }\href
  {https://doi.org/10.1103/PhysRevLett.103.016402} {\bibfield  {journal}
  {\bibinfo  {journal} {Phys. Rev. Lett.}\ }\textbf {\bibinfo {volume} {103}},\
  \bibinfo {pages} {016402} (\bibinfo {year} {2009})}\BibitemShut {NoStop}%
\bibitem [{\citenamefont {Montambaux}\ \emph {et~al.}(2009)\citenamefont
  {Montambaux}, \citenamefont {Piéchon}, \citenamefont {Fuchs},\ and\
  \citenamefont {Goerbig}}]{Montambaux2009}%
  \BibitemOpen
  \bibfield  {author} {\bibinfo {author} {\bibfnamefont {G.}~\bibnamefont
  {Montambaux}}, \bibinfo {author} {\bibfnamefont {F.}~\bibnamefont
  {Piéchon}}, \bibinfo {author} {\bibfnamefont {J.-N.}\ \bibnamefont
  {Fuchs}},\ and\ \bibinfo {author} {\bibfnamefont {M.~O.}\ \bibnamefont
  {Goerbig}},\ }\bibfield  {title} {\bibinfo {title} {Merging of dirac points
  in a two-dimensional crystal},\ }\href
  {https://doi.org/10.1103/PhysRevB.80.153412} {\bibfield  {journal} {\bibinfo
  {journal} {Phys. Rev. B}\ }\textbf {\bibinfo {volume} {80}},\ \bibinfo
  {pages} {153412} (\bibinfo {year} {2009})}\BibitemShut {NoStop}%
\bibitem [{\citenamefont {Delplace}\ and\ \citenamefont
  {Montambaux}(2010)}]{Delplace2010}%
  \BibitemOpen
  \bibfield  {author} {\bibinfo {author} {\bibfnamefont {P.}~\bibnamefont
  {Delplace}}\ and\ \bibinfo {author} {\bibfnamefont {G.}~\bibnamefont
  {Montambaux}},\ }\bibfield  {title} {\bibinfo {title} {Semi-dirac point in
  the hofstadter spectrum},\ }\href
  {https://doi.org/10.1103/PhysRevB.82.035438} {\bibfield  {journal} {\bibinfo
  {journal} {Phys. Rev. B}\ }\textbf {\bibinfo {volume} {82}},\ \bibinfo
  {pages} {035438} (\bibinfo {year} {2010})}\BibitemShut {NoStop}%
\bibitem [{\citenamefont {Dóra}\ \emph {et~al.}(2013)\citenamefont {Dóra},
  \citenamefont {Herbut},\ and\ \citenamefont {Moessner}}]{Dora2013}%
  \BibitemOpen
  \bibfield  {author} {\bibinfo {author} {\bibfnamefont {B.}~\bibnamefont
  {Dóra}}, \bibinfo {author} {\bibfnamefont {I.~F.}\ \bibnamefont {Herbut}},\
  and\ \bibinfo {author} {\bibfnamefont {R.}~\bibnamefont {Moessner}},\
  }\bibfield  {title} {\bibinfo {title} {Coupling, merging, and splitting dirac
  points by electron-electron interaction},\ }\href
  {https://doi.org/10.1103/PhysRevB.88.075126} {\bibfield  {journal} {\bibinfo
  {journal} {Phys. Rev. B}\ }\textbf {\bibinfo {volume} {88}},\ \bibinfo
  {pages} {075126} (\bibinfo {year} {2013})}\BibitemShut {NoStop}%
\bibitem [{\citenamefont {Huang}\ \emph {et~al.}(2015)\citenamefont {Huang},
  \citenamefont {Liu}, \citenamefont {Zhang}, \citenamefont {Duan},\ and\
  \citenamefont {Vanderbilt}}]{Huang2015}%
  \BibitemOpen
  \bibfield  {author} {\bibinfo {author} {\bibfnamefont {H.}~\bibnamefont
  {Huang}}, \bibinfo {author} {\bibfnamefont {Z.}~\bibnamefont {Liu}}, \bibinfo
  {author} {\bibfnamefont {H.}~\bibnamefont {Zhang}}, \bibinfo {author}
  {\bibfnamefont {W.}~\bibnamefont {Duan}},\ and\ \bibinfo {author}
  {\bibfnamefont {D.}~\bibnamefont {Vanderbilt}},\ }\bibfield  {title}
  {\bibinfo {title} {Emergence of a chern-insulating state from a semi-dirac
  dispersion},\ }\href {https://doi.org/10.1103/PhysRevB.92.161115} {\bibfield
  {journal} {\bibinfo  {journal} {Phys. Rev. B}\ }\textbf {\bibinfo {volume}
  {92}},\ \bibinfo {pages} {161115(R)} (\bibinfo {year} {2015})}\BibitemShut
  {NoStop}%
\bibitem [{\citenamefont {Saha}(2016)}]{Saha2016}%
  \BibitemOpen
  \bibfield  {author} {\bibinfo {author} {\bibfnamefont {K.}~\bibnamefont
  {Saha}},\ }\bibfield  {title} {\bibinfo {title} {Photoinduced chern
  insulating states in semi-dirac materials},\ }\href
  {https://doi.org/10.1103/PhysRevB.94.081103} {\bibfield  {journal} {\bibinfo
  {journal} {Phys. Rev. B}\ }\textbf {\bibinfo {volume} {94}},\ \bibinfo
  {pages} {081103(R)} (\bibinfo {year} {2016})}\BibitemShut {NoStop}%
\bibitem [{\citenamefont {Roy}\ and\ \citenamefont {Foster}(2018)}]{Roy2018}%
  \BibitemOpen
  \bibfield  {author} {\bibinfo {author} {\bibfnamefont {B.}~\bibnamefont
  {Roy}}\ and\ \bibinfo {author} {\bibfnamefont {M.~S.}\ \bibnamefont
  {Foster}},\ }\bibfield  {title} {\bibinfo {title} {Quantum multicriticality
  near the dirac-semimetal to band-insulator critical point in two dimensions:
  A controlled ascent from one dimension},\ }\href
  {https://doi.org/10.1103/PhysRevX.8.011049} {\bibfield  {journal} {\bibinfo
  {journal} {Phys. Rev. X}\ }\textbf {\bibinfo {volume} {8}},\ \bibinfo {pages}
  {011049} (\bibinfo {year} {2018})}\BibitemShut {NoStop}%
\bibitem [{\citenamefont {Uryszek}\ \emph {et~al.}(2019)\citenamefont
  {Uryszek}, \citenamefont {Christou}, \citenamefont {Jaefari}, \citenamefont
  {Krüger},\ and\ \citenamefont {Uchoa}}]{Uryszek2019}%
  \BibitemOpen
  \bibfield  {author} {\bibinfo {author} {\bibfnamefont {M.~D.}\ \bibnamefont
  {Uryszek}}, \bibinfo {author} {\bibfnamefont {E.}~\bibnamefont {Christou}},
  \bibinfo {author} {\bibfnamefont {A.}~\bibnamefont {Jaefari}}, \bibinfo
  {author} {\bibfnamefont {F.}~\bibnamefont {Krüger}},\ and\ \bibinfo {author}
  {\bibfnamefont {B.}~\bibnamefont {Uchoa}},\ }\bibfield  {title} {\bibinfo
  {title} {Quantum criticality of semi-dirac fermions in $2+1$ dimensions},\
  }\href {https://doi.org/10.1103/PhysRevB.100.155101} {\bibfield  {journal}
  {\bibinfo  {journal} {Phys. Rev. B}\ }\textbf {\bibinfo {volume} {100}},\
  \bibinfo {pages} {155101} (\bibinfo {year} {2019})}\BibitemShut {NoStop}%
\bibitem [{\citenamefont {Kotov}\ \emph {et~al.}(2021)\citenamefont {Kotov},
  \citenamefont {Uchoa},\ and\ \citenamefont {Sushkov}}]{Kotov2021}%
  \BibitemOpen
  \bibfield  {author} {\bibinfo {author} {\bibfnamefont {V.~N.}\ \bibnamefont
  {Kotov}}, \bibinfo {author} {\bibfnamefont {B.}~\bibnamefont {Uchoa}},\ and\
  \bibinfo {author} {\bibfnamefont {O.~P.}\ \bibnamefont {Sushkov}},\
  }\bibfield  {title} {\bibinfo {title} {Coulomb interactions and
  renormalization of semi-dirac fermions near a topological lifshitz
  transition},\ }\href {https://doi.org/10.1103/PhysRevB.103.045403} {\bibfield
   {journal} {\bibinfo  {journal} {Phys. Rev. B}\ }\textbf {\bibinfo {volume}
  {103}},\ \bibinfo {pages} {045403} (\bibinfo {year} {2021})}\BibitemShut
  {NoStop}%
\bibitem [{\citenamefont {Mohanta}\ \emph {et~al.}(2021)\citenamefont {Mohanta}
  \emph {et~al.}}]{Mohanta2021}%
  \BibitemOpen
  \bibfield  {author} {\bibinfo {author} {\bibfnamefont {N.}~\bibnamefont
  {Mohanta}} \emph {et~al.},\ }\bibfield  {title} {\bibinfo {title} {Semi-dirac
  and weyl fermions in transition metal oxides},\ }\href
  {https://doi.org/10.1103/PhysRevB.104.235121} {\bibfield  {journal} {\bibinfo
   {journal} {Phys. Rev. B}\ }\textbf {\bibinfo {volume} {104}},\ \bibinfo
  {pages} {235121} (\bibinfo {year} {2021})}\BibitemShut {NoStop}%
\bibitem [{\citenamefont {Yuan}\ \emph {et~al.}(2016)\citenamefont {Yuan},
  \citenamefont {Zhang}, \citenamefont {Liu}, \citenamefont {Narayan},
  \citenamefont {Song}, \citenamefont {Shen}, \citenamefont {Sui},
  \citenamefont {Xu}, \citenamefont {Yu}, \citenamefont {An}, \citenamefont
  {Zhao}, \citenamefont {Sanvito}, \citenamefont {Yan},\ and\ \citenamefont
  {Xiu}}]{yuan2016b}%
  \BibitemOpen
  \bibfield  {author} {\bibinfo {author} {\bibfnamefont {X.}~\bibnamefont
  {Yuan}}, \bibinfo {author} {\bibfnamefont {C.}~\bibnamefont {Zhang}},
  \bibinfo {author} {\bibfnamefont {Y.}~\bibnamefont {Liu}}, \bibinfo {author}
  {\bibfnamefont {A.}~\bibnamefont {Narayan}}, \bibinfo {author} {\bibfnamefont
  {C.}~\bibnamefont {Song}}, \bibinfo {author} {\bibfnamefont {S.}~\bibnamefont
  {Shen}}, \bibinfo {author} {\bibfnamefont {X.}~\bibnamefont {Sui}}, \bibinfo
  {author} {\bibfnamefont {J.}~\bibnamefont {Xu}}, \bibinfo {author}
  {\bibfnamefont {H.}~\bibnamefont {Yu}}, \bibinfo {author} {\bibfnamefont
  {Z.}~\bibnamefont {An}}, \bibinfo {author} {\bibfnamefont {J.}~\bibnamefont
  {Zhao}}, \bibinfo {author} {\bibfnamefont {S.}~\bibnamefont {Sanvito}},
  \bibinfo {author} {\bibfnamefont {H.}~\bibnamefont {Yan}},\ and\ \bibinfo
  {author} {\bibfnamefont {F.}~\bibnamefont {Xiu}},\ }\bibfield  {title}
  {\bibinfo {title} {Observation of quasi-two-dimensional {{Dirac}} fermions in
  {ZrTe$_5$}},\ }\href {https://doi.org/10.1038/am.2016.166} {\bibfield
  {journal} {\bibinfo  {journal} {NPG Asia Mater.}\ }\textbf {\bibinfo {volume}
  {8}},\ \bibinfo {pages} {e325} (\bibinfo {year} {2016})}\BibitemShut
  {NoStop}%
\bibitem [{\citenamefont {Schoop}\ \emph {et~al.}(2016)\citenamefont {Schoop}
  \emph {et~al.}}]{Schoop2016}%
  \BibitemOpen
  \bibfield  {author} {\bibinfo {author} {\bibfnamefont {L.~M.}\ \bibnamefont
  {Schoop}} \emph {et~al.},\ }\bibfield  {title} {\bibinfo {title} {Dirac cone
  protected by non-symmorphic symmetry and three-dimensional dirac line node in
  $\rm{ZrSiS}$},\ }\href {https://doi.org/10.1038/ncomms11696} {\bibfield
  {journal} {\bibinfo  {journal} {Nat. Commun.}\ }\textbf {\bibinfo {volume}
  {7}},\ \bibinfo {pages} {11696} (\bibinfo {year} {2016})}\BibitemShut
  {NoStop}%
\bibitem [{\citenamefont {Hu}\ \emph {et~al.}(2016)\citenamefont {Hu} \emph
  {et~al.}}]{Hu2016}%
  \BibitemOpen
  \bibfield  {author} {\bibinfo {author} {\bibfnamefont {J.}~\bibnamefont {Hu}}
  \emph {et~al.},\ }\bibfield  {title} {\bibinfo {title} {Evidence of
  topological nodal-line fermions in $\rm{ZrSiSe}$ and $\rm{ZrSiTe}$},\ }\href
  {https://doi.org/10.1103/PhysRevLett.117.016602} {\bibfield  {journal}
  {\bibinfo  {journal} {Phys. Rev. Lett.}\ }\textbf {\bibinfo {volume} {117}},\
  \bibinfo {pages} {016602} (\bibinfo {year} {2016})}\BibitemShut {NoStop}%
\bibitem [{\citenamefont {Kim}\ \emph {et~al.}(2015{\natexlab{a}})\citenamefont
  {Kim}, \citenamefont {Wieder}, \citenamefont {Kane},\ and\ \citenamefont
  {Rappe}}]{Kim2015}%
  \BibitemOpen
  \bibfield  {author} {\bibinfo {author} {\bibfnamefont {Y.}~\bibnamefont
  {Kim}}, \bibinfo {author} {\bibfnamefont {B.~J.}\ \bibnamefont {Wieder}},
  \bibinfo {author} {\bibfnamefont {C.~L.}\ \bibnamefont {Kane}},\ and\
  \bibinfo {author} {\bibfnamefont {A.~M.}\ \bibnamefont {Rappe}},\ }\bibfield
  {title} {\bibinfo {title} {Dirac line nodes in inversion-symmetric
  crystals},\ }\href {https://doi.org/10.1103/PhysRevLett.115.036806}
  {\bibfield  {journal} {\bibinfo  {journal} {Phys. Rev. Lett.}\ }\textbf
  {\bibinfo {volume} {115}},\ \bibinfo {pages} {036806} (\bibinfo {year}
  {2015}{\natexlab{a}})}\BibitemShut {NoStop}%
\bibitem [{\citenamefont {Fang}\ \emph {et~al.}(2015)\citenamefont {Fang},
  \citenamefont {Chen}, \citenamefont {Kee},\ and\ \citenamefont
  {Fu}}]{Fang2015}%
  \BibitemOpen
  \bibfield  {author} {\bibinfo {author} {\bibfnamefont {C.}~\bibnamefont
  {Fang}}, \bibinfo {author} {\bibfnamefont {Y.}~\bibnamefont {Chen}}, \bibinfo
  {author} {\bibfnamefont {H.-Y.}\ \bibnamefont {Kee}},\ and\ \bibinfo {author}
  {\bibfnamefont {L.}~\bibnamefont {Fu}},\ }\bibfield  {title} {\bibinfo
  {title} {Topological nodal line semimetals with and without spin-orbital
  coupling},\ }\href {https://doi.org/10.1103/PhysRevB.92.081201} {\bibfield
  {journal} {\bibinfo  {journal} {Phys. Rev. B}\ }\textbf {\bibinfo {volume}
  {92}},\ \bibinfo {pages} {081201(R)} (\bibinfo {year} {2015})}\BibitemShut
  {NoStop}%
\bibitem [{\citenamefont {Bzdušek}\ \emph {et~al.}(2016)\citenamefont
  {Bzdušek}, \citenamefont {Wu}, \citenamefont {Rüegg}, \citenamefont
  {Sigrist},\ and\ \citenamefont {Soluyanov}}]{Bzdusek2016}%
  \BibitemOpen
  \bibfield  {author} {\bibinfo {author} {\bibfnamefont {T.}~\bibnamefont
  {Bzdušek}}, \bibinfo {author} {\bibfnamefont {Q.}~\bibnamefont {Wu}},
  \bibinfo {author} {\bibfnamefont {A.}~\bibnamefont {Rüegg}}, \bibinfo
  {author} {\bibfnamefont {M.}~\bibnamefont {Sigrist}},\ and\ \bibinfo {author}
  {\bibfnamefont {A.~A.}\ \bibnamefont {Soluyanov}},\ }\bibfield  {title}
  {\bibinfo {title} {Nodal-chain metals},\ }\href
  {https://doi.org/10.1038/nature19099} {\bibfield  {journal} {\bibinfo
  {journal} {Nature}\ }\textbf {\bibinfo {volume} {538}},\ \bibinfo {pages}
  {75} (\bibinfo {year} {2016})}\BibitemShut {NoStop}%
\bibitem [{\citenamefont {Yan}\ \emph {et~al.}(2018)\citenamefont {Yan} \emph
  {et~al.}}]{Yan2018}%
  \BibitemOpen
  \bibfield  {author} {\bibinfo {author} {\bibfnamefont {Q.}~\bibnamefont
  {Yan}} \emph {et~al.},\ }\bibfield  {title} {\bibinfo {title} {Experimental
  discovery of nodal chains},\ }\href
  {https://doi.org/10.1038/s41567-018-0041-4} {\bibfield  {journal} {\bibinfo
  {journal} {Nat. Phys.}\ }\textbf {\bibinfo {volume} {14}},\ \bibinfo {pages}
  {461–464} (\bibinfo {year} {2018})}\BibitemShut {NoStop}%
\bibitem [{\citenamefont {Chang}\ \emph {et~al.}(2017)\citenamefont {Chang}
  \emph {et~al.}}]{Chang2017}%
  \BibitemOpen
  \bibfield  {author} {\bibinfo {author} {\bibfnamefont {G.}~\bibnamefont
  {Chang}} \emph {et~al.},\ }\bibfield  {title} {\bibinfo {title} {Topological
  hopf and chain link semimetal states and their application to co2mnga},\
  }\href {https://doi.org/10.1103/PhysRevLett.119.156401} {\bibfield  {journal}
  {\bibinfo  {journal} {Phys. Rev. Lett.}\ }\textbf {\bibinfo {volume} {119}},\
  \bibinfo {pages} {156401} (\bibinfo {year} {2017})}\BibitemShut {NoStop}%
\bibitem [{\citenamefont {Chen}\ \emph
  {et~al.}(2017{\natexlab{a}})\citenamefont {Chen}, \citenamefont {Lu},\ and\
  \citenamefont {Hou}}]{Chen2017}%
  \BibitemOpen
  \bibfield  {author} {\bibinfo {author} {\bibfnamefont {W.}~\bibnamefont
  {Chen}}, \bibinfo {author} {\bibfnamefont {H.-Z.}\ \bibnamefont {Lu}},\ and\
  \bibinfo {author} {\bibfnamefont {J.-M.}\ \bibnamefont {Hou}},\ }\bibfield
  {title} {\bibinfo {title} {Topological semimetals with a double-helix nodal
  link},\ }\href {https://doi.org/10.1103/PhysRevB.96.041102} {\bibfield
  {journal} {\bibinfo  {journal} {Phys. Rev. B}\ }\textbf {\bibinfo {volume}
  {96}},\ \bibinfo {pages} {041102(R)} (\bibinfo {year}
  {2017}{\natexlab{a}})}\BibitemShut {NoStop}%
\bibitem [{\citenamefont {Wu}\ \emph {et~al.}(2019)\citenamefont {Wu},
  \citenamefont {Soluyanov},\ and\ \citenamefont {Bzdušek}}]{Wu2019}%
  \BibitemOpen
  \bibfield  {author} {\bibinfo {author} {\bibfnamefont {Q.}~\bibnamefont
  {Wu}}, \bibinfo {author} {\bibfnamefont {A.~A.}\ \bibnamefont {Soluyanov}},\
  and\ \bibinfo {author} {\bibfnamefont {T.}~\bibnamefont {Bzdušek}},\
  }\bibfield  {title} {\bibinfo {title} {Non-abelian band topology in
  noninteracting metals},\ }\href {https://doi.org/10.1126/science.aau8740}
  {\bibfield  {journal} {\bibinfo  {journal} {Science}\ }\textbf {\bibinfo
  {volume} {365}},\ \bibinfo {pages} {1273–1277} (\bibinfo {year}
  {2019})}\BibitemShut {NoStop}%
\bibitem [{\citenamefont {Novoselov}\ \emph {et~al.}(2005)\citenamefont
  {Novoselov} \emph {et~al.}}]{Novoselov2005}%
  \BibitemOpen
  \bibfield  {author} {\bibinfo {author} {\bibfnamefont {K.~S.}\ \bibnamefont
  {Novoselov}} \emph {et~al.},\ }\bibfield  {title} {\bibinfo {title}
  {Two-dimensional gas of massless dirac fermions in graphene},\ }\href
  {https://doi.org/10.1038/nature04233} {\bibfield  {journal} {\bibinfo
  {journal} {Nature}\ }\textbf {\bibinfo {volume} {438}},\ \bibinfo {pages}
  {197–200} (\bibinfo {year} {2005})}\BibitemShut {NoStop}%
\bibitem [{\citenamefont {Zhang}\ \emph {et~al.}(2005)\citenamefont {Zhang},
  \citenamefont {Tan}, \citenamefont {Stormer},\ and\ \citenamefont
  {Kim}}]{Zhang2005}%
  \BibitemOpen
  \bibfield  {author} {\bibinfo {author} {\bibfnamefont {Y.}~\bibnamefont
  {Zhang}}, \bibinfo {author} {\bibfnamefont {Y.-W.}\ \bibnamefont {Tan}},
  \bibinfo {author} {\bibfnamefont {H.~L.}\ \bibnamefont {Stormer}},\ and\
  \bibinfo {author} {\bibfnamefont {P.}~\bibnamefont {Kim}},\ }\bibfield
  {title} {\bibinfo {title} {Experimental observation of the quantum hall
  effect and berry's phase in graphene},\ }\href
  {https://doi.org/10.1038/nature04235} {\bibfield  {journal} {\bibinfo
  {journal} {Nature}\ }\textbf {\bibinfo {volume} {438}},\ \bibinfo {pages}
  {201–204} (\bibinfo {year} {2005})}\BibitemShut {NoStop}%
\bibitem [{\citenamefont {Katsnelson}\ \emph {et~al.}(2006)\citenamefont
  {Katsnelson}, \citenamefont {Novoselov},\ and\ \citenamefont
  {Geim}}]{Katsnelson2006}%
  \BibitemOpen
  \bibfield  {author} {\bibinfo {author} {\bibfnamefont {M.~I.}\ \bibnamefont
  {Katsnelson}}, \bibinfo {author} {\bibfnamefont {K.~S.}\ \bibnamefont
  {Novoselov}},\ and\ \bibinfo {author} {\bibfnamefont {A.~K.}\ \bibnamefont
  {Geim}},\ }\bibfield  {title} {\bibinfo {title} {Chiral tunnelling and the
  klein paradox in graphene},\ }\href {https://doi.org/10.1038/nphys384}
  {\bibfield  {journal} {\bibinfo  {journal} {Nat. Phys.}\ }\textbf {\bibinfo
  {volume} {2}},\ \bibinfo {pages} {620–625} (\bibinfo {year}
  {2006})}\BibitemShut {NoStop}%
\bibitem [{\citenamefont {Young}\ and\ \citenamefont {Kim}(2009)}]{Young2009}%
  \BibitemOpen
  \bibfield  {author} {\bibinfo {author} {\bibfnamefont {A.~F.}\ \bibnamefont
  {Young}}\ and\ \bibinfo {author} {\bibfnamefont {P.}~\bibnamefont {Kim}},\
  }\bibfield  {title} {\bibinfo {title} {Quantum interference and klein
  tunnelling in graphene heterojunctions},\ }\href
  {https://doi.org/10.1038/nphys1198} {\bibfield  {journal} {\bibinfo
  {journal} {Nat. Phys.}\ }\textbf {\bibinfo {volume} {5}},\ \bibinfo {pages}
  {222} (\bibinfo {year} {2009})}\BibitemShut {NoStop}%
\bibitem [{\citenamefont {Sadowski}\ \emph {et~al.}(2006)\citenamefont
  {Sadowski}, \citenamefont {Martinez}, \citenamefont {Potemski}, \citenamefont
  {Berger},\ and\ \citenamefont {de~Heer}}]{Sadowski2006}%
  \BibitemOpen
  \bibfield  {author} {\bibinfo {author} {\bibfnamefont {M.~L.}\ \bibnamefont
  {Sadowski}}, \bibinfo {author} {\bibfnamefont {G.}~\bibnamefont {Martinez}},
  \bibinfo {author} {\bibfnamefont {M.}~\bibnamefont {Potemski}}, \bibinfo
  {author} {\bibfnamefont {C.}~\bibnamefont {Berger}},\ and\ \bibinfo {author}
  {\bibfnamefont {W.~A.}\ \bibnamefont {de~Heer}},\ }\bibfield  {title}
  {\bibinfo {title} {Landau level spectroscopy of ultrathin graphite layers},\
  }\href {https://doi.org/10.1103/PhysRevLett.97.266405} {\bibfield  {journal}
  {\bibinfo  {journal} {Phys. Rev. Lett.}\ }\textbf {\bibinfo {volume} {97}},\
  \bibinfo {pages} {266405} (\bibinfo {year} {2006})}\BibitemShut {NoStop}%
\bibitem [{\citenamefont {Jiang}\ \emph {et~al.}(2007)\citenamefont {Jiang}
  \emph {et~al.}}]{Jiang2007}%
  \BibitemOpen
  \bibfield  {author} {\bibinfo {author} {\bibfnamefont {Z.}~\bibnamefont
  {Jiang}} \emph {et~al.},\ }\bibfield  {title} {\bibinfo {title} {Infrared
  spectroscopy of landau levels of graphene},\ }\href
  {https://doi.org/10.1103/PhysRevLett.98.197403} {\bibfield  {journal}
  {\bibinfo  {journal} {Phys. Rev. Lett.}\ }\textbf {\bibinfo {volume} {98}},\
  \bibinfo {pages} {197403} (\bibinfo {year} {2007})}\BibitemShut {NoStop}%
\bibitem [{\citenamefont {Orlita}\ \emph {et~al.}(2008)\citenamefont {Orlita}
  \emph {et~al.}}]{Orlita2008}%
  \BibitemOpen
  \bibfield  {author} {\bibinfo {author} {\bibfnamefont {M.}~\bibnamefont
  {Orlita}} \emph {et~al.},\ }\bibfield  {title} {\bibinfo {title} {Approaching
  the dirac point in high-mobility multilayer epitaxial graphene},\ }\href
  {https://doi.org/10.1103/PhysRevLett.101.267601} {\bibfield  {journal}
  {\bibinfo  {journal} {Phys. Rev. Lett.}\ }\textbf {\bibinfo {volume} {101}},\
  \bibinfo {pages} {267601} (\bibinfo {year} {2008})}\BibitemShut {NoStop}%
\bibitem [{\citenamefont {Goerbig}\ \emph {et~al.}(2008)\citenamefont
  {Goerbig}, \citenamefont {Fuchs}, \citenamefont {Montambaux},\ and\
  \citenamefont {Piéchon}}]{Goerbig2008}%
  \BibitemOpen
  \bibfield  {author} {\bibinfo {author} {\bibfnamefont {M.~O.}\ \bibnamefont
  {Goerbig}}, \bibinfo {author} {\bibfnamefont {J.-N.}\ \bibnamefont {Fuchs}},
  \bibinfo {author} {\bibfnamefont {G.}~\bibnamefont {Montambaux}},\ and\
  \bibinfo {author} {\bibfnamefont {F.}~\bibnamefont {Piéchon}},\ }\bibfield
  {title} {\bibinfo {title} {Tilted anisotropic dirac cones in quinoid-type
  graphene and $\rm{\alpha-(bedt-ttf)_2i_3}$},\ }\href
  {https://doi.org/10.1103/PhysRevB.78.045415} {\bibfield  {journal} {\bibinfo
  {journal} {Phys. Rev. B}\ }\textbf {\bibinfo {volume} {78}},\ \bibinfo
  {pages} {045415} (\bibinfo {year} {2008})}\BibitemShut {NoStop}%
\bibitem [{\citenamefont {Pereira}\ \emph {et~al.}(2009)\citenamefont
  {Pereira}, \citenamefont {Castro~Neto},\ and\ \citenamefont
  {Peres}}]{Pereira2009}%
  \BibitemOpen
  \bibfield  {author} {\bibinfo {author} {\bibfnamefont {V.~M.}\ \bibnamefont
  {Pereira}}, \bibinfo {author} {\bibfnamefont {A.~H.}\ \bibnamefont
  {Castro~Neto}},\ and\ \bibinfo {author} {\bibfnamefont {N.~M.~R.}\
  \bibnamefont {Peres}},\ }\bibfield  {title} {\bibinfo {title} {Tight-binding
  approach to uniaxial strain in graphene},\ }\href
  {https://doi.org/10.1103/PhysRevB.80.045401} {\bibfield  {journal} {\bibinfo
  {journal} {Phys. Rev. B}\ }\textbf {\bibinfo {volume} {80}},\ \bibinfo
  {pages} {045401} (\bibinfo {year} {2009})}\BibitemShut {NoStop}%
\bibitem [{\citenamefont {Kim}\ \emph {et~al.}(2015{\natexlab{b}})\citenamefont
  {Kim} \emph {et~al.}}]{KimJ2015}%
  \BibitemOpen
  \bibfield  {author} {\bibinfo {author} {\bibfnamefont {J.}~\bibnamefont
  {Kim}} \emph {et~al.},\ }\bibfield  {title} {\bibinfo {title} {Observation of
  tunable band gap and anisotropic dirac semimetal state in black phosphorus},\
  }\href {https://doi.org/10.1126/science.aaa6486} {\bibfield  {journal}
  {\bibinfo  {journal} {Science}\ }\textbf {\bibinfo {volume} {349}},\ \bibinfo
  {pages} {723–726} (\bibinfo {year} {2015}{\natexlab{b}})}\BibitemShut
  {NoStop}%
\bibitem [{\citenamefont {Kim}\ \emph {et~al.}(2017)\citenamefont {Kim} \emph
  {et~al.}}]{KimJ2017}%
  \BibitemOpen
  \bibfield  {author} {\bibinfo {author} {\bibfnamefont {J.}~\bibnamefont
  {Kim}} \emph {et~al.},\ }\bibfield  {title} {\bibinfo {title}
  {Two-dimensional dirac fermions protected by space-time inversion symmetry in
  black phosphorus},\ }\href {https://doi.org/10.1103/PhysRevLett.119.226801}
  {\bibfield  {journal} {\bibinfo  {journal} {Phys. Rev. Lett.}\ }\textbf
  {\bibinfo {volume} {119}},\ \bibinfo {pages} {226801} (\bibinfo {year}
  {2017})}\BibitemShut {NoStop}%
\bibitem [{\citenamefont {Rudenko}\ \emph {et~al.}(2015)\citenamefont
  {Rudenko}, \citenamefont {Yuan},\ and\ \citenamefont
  {Katsnelson}}]{Rudenko2015}%
  \BibitemOpen
  \bibfield  {author} {\bibinfo {author} {\bibfnamefont {A.~N.}\ \bibnamefont
  {Rudenko}}, \bibinfo {author} {\bibfnamefont {S.}~\bibnamefont {Yuan}},\ and\
  \bibinfo {author} {\bibfnamefont {M.~I.}\ \bibnamefont {Katsnelson}},\
  }\bibfield  {title} {\bibinfo {title} {Toward a realistic description of
  multilayer black phosphorus: From gw approximation to large-scale
  tight-binding simulations},\ }\href
  {https://doi.org/10.1103/PhysRevB.92.085419} {\bibfield  {journal} {\bibinfo
  {journal} {Phys. Rev. B}\ }\textbf {\bibinfo {volume} {92}},\ \bibinfo
  {pages} {085419} (\bibinfo {year} {2015})}\BibitemShut {NoStop}%
\bibitem [{\citenamefont {Tarruell}\ \emph {et~al.}(2012)\citenamefont
  {Tarruell} \emph {et~al.}}]{Tarruell2012}%
  \BibitemOpen
  \bibfield  {author} {\bibinfo {author} {\bibfnamefont {L.}~\bibnamefont
  {Tarruell}} \emph {et~al.},\ }\bibfield  {title} {\bibinfo {title} {Creating,
  moving, and merging dirac points with a fermi gas in a tunable honeycomb
  lattice},\ }\href {https://doi.org/10.1038/nature10871} {\bibfield  {journal}
  {\bibinfo  {journal} {Nature}\ }\textbf {\bibinfo {volume} {483}},\ \bibinfo
  {pages} {302–305} (\bibinfo {year} {2012})}\BibitemShut {NoStop}%
\bibitem [{\citenamefont {Bellec}\ \emph {et~al.}(2013)\citenamefont {Bellec},
  \citenamefont {Kuhl}, \citenamefont {Montambaux},\ and\ \citenamefont
  {Mortessagne}}]{Bellec2013}%
  \BibitemOpen
  \bibfield  {author} {\bibinfo {author} {\bibfnamefont {M.}~\bibnamefont
  {Bellec}}, \bibinfo {author} {\bibfnamefont {U.}~\bibnamefont {Kuhl}},
  \bibinfo {author} {\bibfnamefont {G.}~\bibnamefont {Montambaux}},\ and\
  \bibinfo {author} {\bibfnamefont {F.}~\bibnamefont {Mortessagne}},\
  }\bibfield  {title} {\bibinfo {title} {Topological transition of dirac points
  in a microwave experiment},\ }\href
  {https://doi.org/10.1103/PhysRevLett.110.033902} {\bibfield  {journal}
  {\bibinfo  {journal} {Phys. Rev. Lett.}\ }\textbf {\bibinfo {volume} {110}},\
  \bibinfo {pages} {033902} (\bibinfo {year} {2013})}\BibitemShut {NoStop}%
\bibitem [{\citenamefont {Rechtsman}\ \emph {et~al.}(2013)\citenamefont
  {Rechtsman} \emph {et~al.}}]{Rechtsman2013}%
  \BibitemOpen
  \bibfield  {author} {\bibinfo {author} {\bibfnamefont {M.~C.}\ \bibnamefont
  {Rechtsman}} \emph {et~al.},\ }\bibfield  {title} {\bibinfo {title}
  {Topological creation and destruction of edge states in photonic graphene},\
  }\href {https://doi.org/10.1103/PhysRevLett.111.103901} {\bibfield  {journal}
  {\bibinfo  {journal} {Phys. Rev. Lett.}\ }\textbf {\bibinfo {volume} {111}},\
  \bibinfo {pages} {103901} (\bibinfo {year} {2013})}\BibitemShut {NoStop}%
\bibitem [{\citenamefont {Real}\ \emph {et~al.}(2020)\citenamefont {Real} \emph
  {et~al.}}]{Real2020}%
  \BibitemOpen
  \bibfield  {author} {\bibinfo {author} {\bibfnamefont {B.}~\bibnamefont
  {Real}} \emph {et~al.},\ }\bibfield  {title} {\bibinfo {title} {Semi-dirac
  transport and anisotropic localization in polariton honeycomb lattices},\
  }\href {https://doi.org/10.1103/PhysRevLett.125.186601} {\bibfield  {journal}
  {\bibinfo  {journal} {Phys. Rev. Lett.}\ }\textbf {\bibinfo {volume} {125}},\
  \bibinfo {pages} {186601} (\bibinfo {year} {2020})}\BibitemShut {NoStop}%
\bibitem [{\citenamefont {Isobe}\ \emph {et~al.}(2016)\citenamefont {Isobe}
  \emph {et~al.}}]{Isobe2016}%
  \BibitemOpen
  \bibfield  {author} {\bibinfo {author} {\bibfnamefont {H.}~\bibnamefont
  {Isobe}} \emph {et~al.},\ }\bibfield  {title} {\bibinfo {title} {Emergent
  non-fermi-liquid at the quantum critical point of a topological phase
  transition in two dimensions},\ }\href
  {https://doi.org/10.1103/PhysRevLett.116.076803} {\bibfield  {journal}
  {\bibinfo  {journal} {Phys. Rev. Lett.}\ }\textbf {\bibinfo {volume} {116}},\
  \bibinfo {pages} {076803} (\bibinfo {year} {2016})}\BibitemShut {NoStop}%
\bibitem [{\citenamefont {Onsager}(1952)}]{Onsager1952}%
  \BibitemOpen
  \bibfield  {author} {\bibinfo {author} {\bibfnamefont {L.}~\bibnamefont
  {Onsager}},\ }\bibfield  {title} {\bibinfo {title} {Interpretation of the de
  haas-van alphen effect},\ }\href {https://doi.org/10.1080/14786440908521019}
  {\bibfield  {journal} {\bibinfo  {journal} {The London, Edinburgh, and Dublin
  Philosophical Magazine and Journal of Science}\ }\textbf {\bibinfo {volume}
  {43}},\ \bibinfo {pages} {1006–1008} (\bibinfo {year} {1952})}\BibitemShut
  {NoStop}%
\bibitem [{\citenamefont {Muechler}\ \emph {et~al.}(2020)\citenamefont
  {Muechler} \emph {et~al.}}]{Muechler2020}%
  \BibitemOpen
  \bibfield  {author} {\bibinfo {author} {\bibfnamefont {L.}~\bibnamefont
  {Muechler}} \emph {et~al.},\ }\bibfield  {title} {\bibinfo {title} {Modular
  arithmetic with nodal lines: Drumhead surface states in $\rm{ZrSiTe}$},\
  }\href {https://doi.org/10.1103/PhysRevX.10.011026} {\bibfield  {journal}
  {\bibinfo  {journal} {Phys. Rev. X}\ }\textbf {\bibinfo {volume} {10}},\
  \bibinfo {pages} {011026} (\bibinfo {year} {2020})}\BibitemShut {NoStop}%
\bibitem [{\citenamefont {Schilling}\ \emph {et~al.}(2017)\citenamefont
  {Schilling}, \citenamefont {Schoop}, \citenamefont {Lotsch}, \citenamefont
  {Dressel},\ and\ \citenamefont {Pronin}}]{Schilling2017}%
  \BibitemOpen
  \bibfield  {author} {\bibinfo {author} {\bibfnamefont {M.~B.}\ \bibnamefont
  {Schilling}}, \bibinfo {author} {\bibfnamefont {L.~M.}\ \bibnamefont
  {Schoop}}, \bibinfo {author} {\bibfnamefont {B.~V.}\ \bibnamefont {Lotsch}},
  \bibinfo {author} {\bibfnamefont {M.}~\bibnamefont {Dressel}},\ and\ \bibinfo
  {author} {\bibfnamefont {A.~V.}\ \bibnamefont {Pronin}},\ }\bibfield  {title}
  {\bibinfo {title} {Flat optical conductivity in zrsis due to two-dimensional
  dirac bands},\ }\href {https://doi.org/10.1103/PhysRevLett.119.187401}
  {\bibfield  {journal} {\bibinfo  {journal} {Phys. Rev. Lett.}\ }\textbf
  {\bibinfo {volume} {119}},\ \bibinfo {pages} {187401} (\bibinfo {year}
  {2017})}\BibitemShut {NoStop}%
\bibitem [{\citenamefont {Rudenko}\ \emph {et~al.}(2018)\citenamefont
  {Rudenko}, \citenamefont {Stepanov}, \citenamefont {Lichtenstein},\ and\
  \citenamefont {Katsnelson}}]{Rudenko2018}%
  \BibitemOpen
  \bibfield  {author} {\bibinfo {author} {\bibfnamefont {A.~N.}\ \bibnamefont
  {Rudenko}}, \bibinfo {author} {\bibfnamefont {E.~A.}\ \bibnamefont
  {Stepanov}}, \bibinfo {author} {\bibfnamefont {A.~I.}\ \bibnamefont
  {Lichtenstein}},\ and\ \bibinfo {author} {\bibfnamefont {M.~I.}\ \bibnamefont
  {Katsnelson}},\ }\bibfield  {title} {\bibinfo {title} {Excitonic instability
  and pseudogap formation in nodal line semimetal zrsis},\ }\href
  {https://doi.org/10.1103/PhysRevLett.120.216401} {\bibfield  {journal}
  {\bibinfo  {journal} {Phys. Rev. Lett.}\ }\textbf {\bibinfo {volume} {120}},\
  \bibinfo {pages} {216401} (\bibinfo {year} {2018})}\BibitemShut {NoStop}%
\bibitem [{\citenamefont {Shao}\ \emph {et~al.}(2020)\citenamefont {Shao},
  \citenamefont {Rudenko}, \citenamefont {Hu}, \citenamefont {Sun},
  \citenamefont {Zhu}, \citenamefont {Moon}, \citenamefont {Millis},
  \citenamefont {Yuan}, \citenamefont {Lichtenstein}, \citenamefont {Smirnov},
  \citenamefont {Mao}, \citenamefont {Katsnelson},\ and\ \citenamefont
  {Basov}}]{Shao2020}%
  \BibitemOpen
  \bibfield  {author} {\bibinfo {author} {\bibfnamefont {Y.}~\bibnamefont
  {Shao}}, \bibinfo {author} {\bibfnamefont {A.~N.}\ \bibnamefont {Rudenko}},
  \bibinfo {author} {\bibfnamefont {J.}~\bibnamefont {Hu}}, \bibinfo {author}
  {\bibfnamefont {Z.}~\bibnamefont {Sun}}, \bibinfo {author} {\bibfnamefont
  {Y.}~\bibnamefont {Zhu}}, \bibinfo {author} {\bibfnamefont {S.}~\bibnamefont
  {Moon}}, \bibinfo {author} {\bibfnamefont {A.}~\bibnamefont {Millis}},
  \bibinfo {author} {\bibfnamefont {S.}~\bibnamefont {Yuan}}, \bibinfo {author}
  {\bibfnamefont {A.~I.}\ \bibnamefont {Lichtenstein}}, \bibinfo {author}
  {\bibfnamefont {D.}~\bibnamefont {Smirnov}}, \bibinfo {author} {\bibfnamefont
  {Z.~Q.}\ \bibnamefont {Mao}}, \bibinfo {author} {\bibfnamefont {M.~I.}\
  \bibnamefont {Katsnelson}},\ and\ \bibinfo {author} {\bibfnamefont {D.~N.}\
  \bibnamefont {Basov}},\ }\bibfield  {title} {\bibinfo {title} {Electronic
  correlations in nodal-line semimetals},\ }\href
  {https://doi.org/10.1038/s41567-020-0871-0} {\bibfield  {journal} {\bibinfo
  {journal} {Nat. Phys.}\ }\textbf {\bibinfo {volume} {16}},\ \bibinfo {pages}
  {636–641} (\bibinfo {year} {2020})}\BibitemShut {NoStop}%
\bibitem [{\citenamefont {Zhou}\ \emph {et~al.}(2017)\citenamefont {Zhou},
  \citenamefont {Gao}, \citenamefont {Zhang}, \citenamefont {Fang},
  \citenamefont {Song}, \citenamefont {Hu}, \citenamefont {Stroppa},
  \citenamefont {Li}, \citenamefont {Wang}, \citenamefont {Ruan},\ and\
  \citenamefont {Ren}}]{zhou2017}%
  \BibitemOpen
  \bibfield  {author} {\bibinfo {author} {\bibfnamefont {W.}~\bibnamefont
  {Zhou}}, \bibinfo {author} {\bibfnamefont {H.}~\bibnamefont {Gao}}, \bibinfo
  {author} {\bibfnamefont {J.}~\bibnamefont {Zhang}}, \bibinfo {author}
  {\bibfnamefont {R.}~\bibnamefont {Fang}}, \bibinfo {author} {\bibfnamefont
  {H.}~\bibnamefont {Song}}, \bibinfo {author} {\bibfnamefont {T.}~\bibnamefont
  {Hu}}, \bibinfo {author} {\bibfnamefont {A.}~\bibnamefont {Stroppa}},
  \bibinfo {author} {\bibfnamefont {L.}~\bibnamefont {Li}}, \bibinfo {author}
  {\bibfnamefont {X.}~\bibnamefont {Wang}}, \bibinfo {author} {\bibfnamefont
  {S.}~\bibnamefont {Ruan}},\ and\ \bibinfo {author} {\bibfnamefont
  {W.}~\bibnamefont {Ren}},\ }\bibfield  {title} {\bibinfo {title} {Lattice
  dynamics of dirac node-line semimetal zrsis},\ }\href
  {https://doi.org/10.1103/PhysRevB.96.064103} {\bibfield  {journal} {\bibinfo
  {journal} {Phys. Rev. B}\ }\textbf {\bibinfo {volume} {96}},\ \bibinfo
  {pages} {064103} (\bibinfo {year} {2017})}\BibitemShut {NoStop}%
\bibitem [{\citenamefont {Mohelsk\'y}\ \emph {et~al.}(2020)\citenamefont
  {Mohelsk\'y}, \citenamefont {Dubroka}, \citenamefont {Wyzula}, \citenamefont
  {Slobodeniuk}, \citenamefont {Martinez}, \citenamefont {Krupko},
  \citenamefont {Piot}, \citenamefont {Caha}, \citenamefont
  {Huml\'{\i}\ifmmode~\check{c}\else \v{c}\fi{}ek}, \citenamefont {Bauer},
  \citenamefont {Springholz},\ and\ \citenamefont {Orlita}}]{mohelsky2020}%
  \BibitemOpen
  \bibfield  {author} {\bibinfo {author} {\bibfnamefont {I.}~\bibnamefont
  {Mohelsk\'y}}, \bibinfo {author} {\bibfnamefont {A.}~\bibnamefont {Dubroka}},
  \bibinfo {author} {\bibfnamefont {J.}~\bibnamefont {Wyzula}}, \bibinfo
  {author} {\bibfnamefont {A.}~\bibnamefont {Slobodeniuk}}, \bibinfo {author}
  {\bibfnamefont {G.}~\bibnamefont {Martinez}}, \bibinfo {author}
  {\bibfnamefont {Y.}~\bibnamefont {Krupko}}, \bibinfo {author} {\bibfnamefont
  {B.~A.}\ \bibnamefont {Piot}}, \bibinfo {author} {\bibfnamefont
  {O.}~\bibnamefont {Caha}}, \bibinfo {author} {\bibfnamefont {J.}~\bibnamefont
  {Huml\'{\i}\ifmmode~\check{c}\else \v{c}\fi{}ek}}, \bibinfo {author}
  {\bibfnamefont {G.}~\bibnamefont {Bauer}}, \bibinfo {author} {\bibfnamefont
  {G.}~\bibnamefont {Springholz}},\ and\ \bibinfo {author} {\bibfnamefont
  {M.}~\bibnamefont {Orlita}},\ }\bibfield  {title} {\bibinfo {title} {Landau
  level spectroscopy of $\rm{Bi_2Te_3}$},\ }\href
  {https://doi.org/10.1103/PhysRevB.102.085201} {\bibfield  {journal} {\bibinfo
   {journal} {Phys. Rev. B}\ }\textbf {\bibinfo {volume} {102}},\ \bibinfo
  {pages} {085201} (\bibinfo {year} {2020})}\BibitemShut {NoStop}%
\bibitem [{\citenamefont {Polatkan}\ \emph {et~al.}(2023)\citenamefont
  {Polatkan}, \citenamefont {Uykur}, \citenamefont {Mohelsky}, \citenamefont
  {Wyzula}, \citenamefont {Orlita}, \citenamefont {Shekhar}, \citenamefont
  {Felser}, \citenamefont {Dressel},\ and\ \citenamefont
  {Pronin}}]{polatkan2023}%
  \BibitemOpen
  \bibfield  {author} {\bibinfo {author} {\bibfnamefont {S.}~\bibnamefont
  {Polatkan}}, \bibinfo {author} {\bibfnamefont {E.}~\bibnamefont {Uykur}},
  \bibinfo {author} {\bibfnamefont {I.}~\bibnamefont {Mohelsky}}, \bibinfo
  {author} {\bibfnamefont {J.}~\bibnamefont {Wyzula}}, \bibinfo {author}
  {\bibfnamefont {M.}~\bibnamefont {Orlita}}, \bibinfo {author} {\bibfnamefont
  {C.}~\bibnamefont {Shekhar}}, \bibinfo {author} {\bibfnamefont
  {C.}~\bibnamefont {Felser}}, \bibinfo {author} {\bibfnamefont
  {M.}~\bibnamefont {Dressel}},\ and\ \bibinfo {author} {\bibfnamefont {A.~V.}\
  \bibnamefont {Pronin}},\ }\bibfield  {title} {\bibinfo {title} {Exchange gap
  in gdptbi probed by magneto-optics},\ }\href
  {https://doi.org/10.1103/PhysRevB.108.L201201} {\bibfield  {journal}
  {\bibinfo  {journal} {Phys. Rev. B}\ }\textbf {\bibinfo {volume} {108}},\
  \bibinfo {pages} {L201201} (\bibinfo {year} {2023})}\BibitemShut {NoStop}%
\bibitem [{\citenamefont {Shao}\ \emph {et~al.}(2019)\citenamefont {Shao} \emph
  {et~al.}}]{Shao2019}%
  \BibitemOpen
  \bibfield  {author} {\bibinfo {author} {\bibfnamefont {Y.}~\bibnamefont
  {Shao}} \emph {et~al.},\ }\bibfield  {title} {\bibinfo {title} {Optical
  signatures of dirac nodal lines in $\rm{NbAs_2}$},\ }\href
  {https://doi.org/10.1073/pnas.1812821116} {\bibfield  {journal} {\bibinfo
  {journal} {Proc. Natl. Acad. Sci. U.S.A.}\ }\textbf {\bibinfo {volume}
  {116}},\ \bibinfo {pages} {1168} (\bibinfo {year} {2019})}\BibitemShut
  {NoStop}%
\bibitem [{\citenamefont {Santos-Cottin}\ \emph {et~al.}(2021)\citenamefont
  {Santos-Cottin} \emph {et~al.}}]{SantosCottin2021}%
  \BibitemOpen
  \bibfield  {author} {\bibinfo {author} {\bibfnamefont {D.}~\bibnamefont
  {Santos-Cottin}} \emph {et~al.},\ }\bibfield  {title} {\bibinfo {title}
  {Optical conductivity signatures of open dirac nodal lines},\ }\href
  {https://doi.org/10.1103/PhysRevB.104.L201115} {\bibfield  {journal}
  {\bibinfo  {journal} {Phys. Rev. B}\ }\textbf {\bibinfo {volume} {104}},\
  \bibinfo {pages} {L201115} (\bibinfo {year} {2021})}\BibitemShut {NoStop}%
\bibitem [{\citenamefont {Wyzula}\ \emph {et~al.}(2022)\citenamefont {Wyzula}
  \emph {et~al.}}]{Wyzula2022}%
  \BibitemOpen
  \bibfield  {author} {\bibinfo {author} {\bibfnamefont {J.}~\bibnamefont
  {Wyzula}} \emph {et~al.},\ }\bibfield  {title} {\bibinfo {title}
  {Lorentz-boost-driven magneto-optics in a dirac nodal-line semimetal},\
  }\href {https://doi.org/10.1002/advs.202105720} {\bibfield  {journal}
  {\bibinfo  {journal} {Adv. Sci.}\ }\textbf {\bibinfo {volume} {9}},\ \bibinfo
  {pages} {2105720} (\bibinfo {year} {2022})}\BibitemShut {NoStop}%
\bibitem [{\citenamefont {Uykur}\ \emph {et~al.}(2019)\citenamefont {Uykur}
  \emph {et~al.}}]{Uykur2019}%
  \BibitemOpen
  \bibfield  {author} {\bibinfo {author} {\bibfnamefont {E.}~\bibnamefont
  {Uykur}} \emph {et~al.},\ }\bibfield  {title} {\bibinfo {title}
  {Magneto-optical probe of the fully gapped dirac band in $\rm{ZrSiS}$},\
  }\href {https://doi.org/10.1103/PhysRevResearch.1.032015} {\bibfield
  {journal} {\bibinfo  {journal} {Phys. Rev. Research}\ }\textbf {\bibinfo
  {volume} {1}},\ \bibinfo {pages} {032015(R)} (\bibinfo {year}
  {2019})}\BibitemShut {NoStop}%
\bibitem [{\citenamefont {Chen}\ \emph
  {et~al.}(2017{\natexlab{b}})\citenamefont {Chen} \emph {et~al.}}]{chen2017c}%
  \BibitemOpen
  \bibfield  {author} {\bibinfo {author} {\bibfnamefont {Z.-G.}\ \bibnamefont
  {Chen}} \emph {et~al.},\ }\bibfield  {title} {\bibinfo {title} {Spectroscopic
  evidence for bulk-band inversion and three-dimensional massive {{Dirac}}
  fermions in $\rm{ZrTe_5}$},\ }\href {https://doi.org/10.1073/pnas.1613110114}
  {\bibfield  {journal} {\bibinfo  {journal} {Proc. Natl. Acad. Sci. U.S.A.}\
  }\textbf {\bibinfo {volume} {114}},\ \bibinfo {pages} {816} (\bibinfo {year}
  {2017}{\natexlab{b}})}\BibitemShut {NoStop}%
\bibitem [{\citenamefont {Mohelsky}\ \emph {et~al.}(2023)\citenamefont
  {Mohelsky} \emph {et~al.}}]{Mohelsky2023}%
  \BibitemOpen
  \bibfield  {author} {\bibinfo {author} {\bibfnamefont {I.}~\bibnamefont
  {Mohelsky}} \emph {et~al.},\ }\bibfield  {title} {\bibinfo {title}
  {Temperature dependence of the energy band gap in $\rm{ZrTe_5}$: Implications
  for the topological phase},\ }\href
  {https://doi.org/10.1103/PhysRevB.107.L041202} {\bibfield  {journal}
  {\bibinfo  {journal} {Phys. Rev. B}\ }\textbf {\bibinfo {volume} {107}},\
  \bibinfo {pages} {L041202} (\bibinfo {year} {2023})}\BibitemShut {NoStop}%
\bibitem [{\citenamefont {Chen}\ \emph {et~al.}(2015)\citenamefont {Chen} \emph
  {et~al.}}]{Chen2015}%
  \BibitemOpen
  \bibfield  {author} {\bibinfo {author} {\bibfnamefont {R.~Y.}\ \bibnamefont
  {Chen}} \emph {et~al.},\ }\bibfield  {title} {\bibinfo {title}
  {Magnetoinfrared spectroscopy of landau levels and zeeman splitting of
  three-dimensional massless dirac fermions in $\rm{ZrTe_5}$},\ }\href
  {https://doi.org/10.1103/PhysRevLett.115.176404} {\bibfield  {journal}
  {\bibinfo  {journal} {Phys. Rev. Lett.}\ }\textbf {\bibinfo {volume} {115}},\
  \bibinfo {pages} {176404} (\bibinfo {year} {2015})}\BibitemShut {NoStop}%
\bibitem [{\citenamefont {Pack}\ \emph {et~al.}(2020)\citenamefont {Pack} \emph
  {et~al.}}]{pack2020}%
  \BibitemOpen
  \bibfield  {author} {\bibinfo {author} {\bibfnamefont {J.}~\bibnamefont
  {Pack}} \emph {et~al.},\ }\bibfield  {title} {\bibinfo {title} {Broken
  symmetries and kohn’s theorem in graphene cyclotron resonance},\ }\href
  {https://doi.org/10.1103/PhysRevX.10.041006} {\bibfield  {journal} {\bibinfo
  {journal} {Phys. Rev. X}\ }\textbf {\bibinfo {volume} {10}},\ \bibinfo
  {pages} {041006} (\bibinfo {year} {2020})}\BibitemShut {NoStop}%
\bibitem [{\citenamefont {Klemenz}\ \emph {et~al.}(2020)\citenamefont
  {Klemenz}, \citenamefont {Schoop},\ and\ \citenamefont
  {Cano}}]{Klemenz_2020}%
  \BibitemOpen
  \bibfield  {author} {\bibinfo {author} {\bibfnamefont {S.}~\bibnamefont
  {Klemenz}}, \bibinfo {author} {\bibfnamefont {L.}~\bibnamefont {Schoop}},\
  and\ \bibinfo {author} {\bibfnamefont {J.}~\bibnamefont {Cano}},\ }\bibfield
  {title} {\bibinfo {title} {Systematic study of stacked square nets: From
  dirac fermions to material realizations},\ }\href
  {https://doi.org/10.1103/physrevb.101.165121} {\bibfield  {journal} {\bibinfo
   {journal} {Physical Review B}\ }\textbf {\bibinfo {volume} {101}},\ \bibinfo
  {pages} {165121} (\bibinfo {year} {2020})}\BibitemShut {NoStop}%
\bibitem [{\citenamefont {Cao}\ \emph {et~al.}(2018)\citenamefont {Cao} \emph
  {et~al.}}]{Cao2018}%
  \BibitemOpen
  \bibfield  {author} {\bibinfo {author} {\bibfnamefont {Y.}~\bibnamefont
  {Cao}} \emph {et~al.},\ }\bibfield  {title} {\bibinfo {title} {Correlated
  insulator behaviour at half-filling in magic-angle graphene superlattices},\
  }\href {https://doi.org/10.1038/nature26154} {\bibfield  {journal} {\bibinfo
  {journal} {Nature}\ }\textbf {\bibinfo {volume} {556}},\ \bibinfo {pages}
  {80–84} (\bibinfo {year} {2018})}\BibitemShut {NoStop}%
\bibitem [{\citenamefont {Zhou}\ \emph {et~al.}(2021)\citenamefont {Zhou},
  \citenamefont {Chen},\ and\ \citenamefont {Zhu}}]{zhou2021}%
  \BibitemOpen
  \bibfield  {author} {\bibinfo {author} {\bibfnamefont {X.}~\bibnamefont
  {Zhou}}, \bibinfo {author} {\bibfnamefont {W.}~\bibnamefont {Chen}},\ and\
  \bibinfo {author} {\bibfnamefont {X.}~\bibnamefont {Zhu}},\ }\bibfield
  {title} {\bibinfo {title} {Anisotropic magneto-optical absorption and linear
  dichroism in two-dimensional semi-dirac electron systems},\ }\href
  {https://doi.org/10.1103/PhysRevB.104.235403} {\bibfield  {journal} {\bibinfo
   {journal} {Phys. Rev. B}\ }\textbf {\bibinfo {volume} {104}},\ \bibinfo
  {pages} {235403} (\bibinfo {year} {2021})}\BibitemShut {NoStop}%
\bibitem [{\citenamefont {Ahn}\ \emph {et~al.}(2020)\citenamefont {Ahn},
  \citenamefont {Guo},\ and\ \citenamefont {Nagaosa}}]{Ahn2020}%
  \BibitemOpen
  \bibfield  {author} {\bibinfo {author} {\bibfnamefont {J.}~\bibnamefont
  {Ahn}}, \bibinfo {author} {\bibfnamefont {G.-Y.}\ \bibnamefont {Guo}},\ and\
  \bibinfo {author} {\bibfnamefont {N.}~\bibnamefont {Nagaosa}},\ }\bibfield
  {title} {\bibinfo {title} {Low-frequency divergence and quantum geometry of
  the bulk photovoltaic effect in topological semimetals},\ }\href
  {https://doi.org/10.1103/PhysRevX.10.041041} {\bibfield  {journal} {\bibinfo
  {journal} {Phys. Rev. X}\ }\textbf {\bibinfo {volume} {10}},\ \bibinfo
  {pages} {041041} (\bibinfo {year} {2020})}\BibitemShut {NoStop}%
\bibitem [{\citenamefont {Ahn}\ \emph {et~al.}(2022)\citenamefont {Ahn},
  \citenamefont {Guo}, \citenamefont {Nagaosa},\ and\ \citenamefont
  {Vishwanath}}]{Ahn2022}%
  \BibitemOpen
  \bibfield  {author} {\bibinfo {author} {\bibfnamefont {J.}~\bibnamefont
  {Ahn}}, \bibinfo {author} {\bibfnamefont {G.-Y.}\ \bibnamefont {Guo}},
  \bibinfo {author} {\bibfnamefont {N.}~\bibnamefont {Nagaosa}},\ and\ \bibinfo
  {author} {\bibfnamefont {A.}~\bibnamefont {Vishwanath}},\ }\bibfield  {title}
  {\bibinfo {title} {Riemannian geometry of resonant optical responses},\
  }\href {https://doi.org/10.1038/s41567-021-01465-z} {\bibfield  {journal}
  {\bibinfo  {journal} {Nat. Phys.}\ }\textbf {\bibinfo {volume} {18}},\
  \bibinfo {pages} {290–295} (\bibinfo {year} {2022})}\BibitemShut {NoStop}%
\bibitem [{\citenamefont {Onishi}\ and\ \citenamefont
  {Fu}(2024)}]{onishi2024a}%
  \BibitemOpen
  \bibfield  {author} {\bibinfo {author} {\bibfnamefont {Y.}~\bibnamefont
  {Onishi}}\ and\ \bibinfo {author} {\bibfnamefont {L.}~\bibnamefont {Fu}},\
  }\bibfield  {title} {\bibinfo {title} {Fundamental {{Bound}} on {{Topological
  Gap}}},\ }\href {https://doi.org/10.1103/PhysRevX.14.011052} {\bibfield
  {journal} {\bibinfo  {journal} {Phys. Rev. X}\ }\textbf {\bibinfo {volume}
  {14}},\ \bibinfo {pages} {011052} (\bibinfo {year} {2024})}\BibitemShut
  {NoStop}%
\bibitem [{\citenamefont {Souza}\ \emph {et~al.}(2000)\citenamefont {Souza},
  \citenamefont {Wilkens},\ and\ \citenamefont {Martin}}]{Souza2000}%
  \BibitemOpen
  \bibfield  {author} {\bibinfo {author} {\bibfnamefont {I.}~\bibnamefont
  {Souza}}, \bibinfo {author} {\bibfnamefont {T.}~\bibnamefont {Wilkens}},\
  and\ \bibinfo {author} {\bibfnamefont {R.~M.}\ \bibnamefont {Martin}},\
  }\bibfield  {title} {\bibinfo {title} {Polarization and localization in
  insulators: Generating function approach},\ }\href
  {https://doi.org/10.1103/PhysRevB.62.1666} {\bibfield  {journal} {\bibinfo
  {journal} {Phys. Rev. B}\ }\textbf {\bibinfo {volume} {62}},\ \bibinfo
  {pages} {1666} (\bibinfo {year} {2000})}\BibitemShut {NoStop}%
\bibitem [{\citenamefont {Resta}(2011)}]{Resta2011}%
  \BibitemOpen
  \bibfield  {author} {\bibinfo {author} {\bibfnamefont {R.}~\bibnamefont
  {Resta}},\ }\bibfield  {title} {\bibinfo {title} {The insulating state of
  matter: a geometrical theory},\ }\href
  {https://doi.org/10.1140/epjb/e2010-10874-4} {\bibfield  {journal} {\bibinfo
  {journal} {Eur. Phys. J. B}\ }\textbf {\bibinfo {volume} {79}},\ \bibinfo
  {pages} {121–137} (\bibinfo {year} {2011})}\BibitemShut {NoStop}%
\bibitem [{\citenamefont {Kishigi}\ and\ \citenamefont
  {Hasegawa}(2017)}]{Kishigi2017}%
  \BibitemOpen
  \bibfield  {author} {\bibinfo {author} {\bibfnamefont {K.}~\bibnamefont
  {Kishigi}}\ and\ \bibinfo {author} {\bibfnamefont {Y.}~\bibnamefont
  {Hasegawa}},\ }\bibfield  {title} {\bibinfo {title} {Three-quarter dirac
  points, landau levels, and magnetization in $\rm{\alpha-(bedt-ttf)_2i_3}$},\
  }\href {https://doi.org/10.1103/PhysRevB.96.085430} {\bibfield  {journal}
  {\bibinfo  {journal} {Phys. Rev. B}\ }\textbf {\bibinfo {volume} {96}},\
  \bibinfo {pages} {085430} (\bibinfo {year} {2017})}\BibitemShut {NoStop}%
\bibitem [{\citenamefont {Pezzini}\ \emph {et~al.}(2018)\citenamefont {Pezzini}
  \emph {et~al.}}]{Pezzini2018}%
  \BibitemOpen
  \bibfield  {author} {\bibinfo {author} {\bibfnamefont {S.}~\bibnamefont
  {Pezzini}} \emph {et~al.},\ }\bibfield  {title} {\bibinfo {title}
  {Unconventional mass enhancement around the dirac nodal loop in
  $\rm{ZrSiS}$},\ }\href {https://doi.org/10.1038/nphys4306} {\bibfield
  {journal} {\bibinfo  {journal} {Nat. Phys.}\ }\textbf {\bibinfo {volume}
  {14}},\ \bibinfo {pages} {178–183} (\bibinfo {year} {2018})}\BibitemShut
  {NoStop}%
\bibitem [{\citenamefont {Muller}\ \emph {et~al.}(2020)\citenamefont {Muller}
  \emph {et~al.}}]{Muller2020}%
  \BibitemOpen
  \bibfield  {author} {\bibinfo {author} {\bibfnamefont {C.~S.~A.}\
  \bibnamefont {Muller}} \emph {et~al.},\ }\bibfield  {title} {\bibinfo {title}
  {Determination of the fermi surface and field-induced quasiparticle tunneling
  around the dirac nodal loop in $\rm{ZrSiS}$},\ }\href
  {https://doi.org/10.1103/PhysRevResearch.2.023217} {\bibfield  {journal}
  {\bibinfo  {journal} {Phys. Rev. Research}\ }\textbf {\bibinfo {volume}
  {2}},\ \bibinfo {pages} {023217} (\bibinfo {year} {2020})}\BibitemShut
  {NoStop}%
\bibitem [{\citenamefont {Gudac}\ \emph {et~al.}(2022)\citenamefont {Gudac}
  \emph {et~al.}}]{Gudac2022}%
  \BibitemOpen
  \bibfield  {author} {\bibinfo {author} {\bibfnamefont {B.}~\bibnamefont
  {Gudac}} \emph {et~al.},\ }\bibfield  {title} {\bibinfo {title} {Nodal-line
  driven anomalous susceptibility in $\rm{ZrSiS}$},\ }\href
  {https://doi.org/10.1103/PhysRevB.105.L241115} {\bibfield  {journal}
  {\bibinfo  {journal} {Phys. Rev. B}\ }\textbf {\bibinfo {volume} {105}},\
  \bibinfo {pages} {L241115} (\bibinfo {year} {2022})}\BibitemShut {NoStop}%
\bibitem [{\citenamefont {Topp}\ \emph {et~al.}(2017)\citenamefont {Topp} \emph
  {et~al.}}]{Topp2017}%
  \BibitemOpen
  \bibfield  {author} {\bibinfo {author} {\bibfnamefont {A.}~\bibnamefont
  {Topp}} \emph {et~al.},\ }\bibfield  {title} {\bibinfo {title} {Surface
  floating 2d bands in layered nonsymmorphic semimetals: $\rm{ZrSiS}$ and
  related compounds},\ }\href {https://doi.org/10.1103/PhysRevX.7.041073}
  {\bibfield  {journal} {\bibinfo  {journal} {Phys. Rev. X}\ }\textbf {\bibinfo
  {volume} {7}},\ \bibinfo {pages} {041073} (\bibinfo {year}
  {2017})}\BibitemShut {NoStop}%
\bibitem [{\citenamefont {Nakamura}\ \emph {et~al.}(2019)\citenamefont
  {Nakamura} \emph {et~al.}}]{Nakamura2019}%
  \BibitemOpen
  \bibfield  {author} {\bibinfo {author} {\bibfnamefont {T.}~\bibnamefont
  {Nakamura}} \emph {et~al.},\ }\bibfield  {title} {\bibinfo {title} {Evidence
  for bulk nodal loops and universality of dirac-node arc surface states in
  $\rm{ZrGeXc}$ (xc = s, se, te)},\ }\href
  {https://doi.org/10.1103/PhysRevB.99.245105} {\bibfield  {journal} {\bibinfo
  {journal} {Phys. Rev. B}\ }\textbf {\bibinfo {volume} {99}},\ \bibinfo
  {pages} {245105} (\bibinfo {year} {2019})}\BibitemShut {NoStop}%
\bibitem [{\citenamefont {Giannozzi}\ \emph {et~al.}(2009)\citenamefont
  {Giannozzi} \emph {et~al.}}]{Giannozzi2009}%
  \BibitemOpen
  \bibfield  {author} {\bibinfo {author} {\bibfnamefont {P.}~\bibnamefont
  {Giannozzi}} \emph {et~al.},\ }\bibfield  {title} {\bibinfo {title} {Quantum
  espresso: a modular and open-source software project for quantum simulations
  of materials},\ }\href {https://doi.org/10.1088/0953-8984/21/39/395502}
  {\bibfield  {journal} {\bibinfo  {journal} {J. Phys.: Condens. Matter}\
  }\textbf {\bibinfo {volume} {21}},\ \bibinfo {pages} {395502} (\bibinfo
  {year} {2009})}\BibitemShut {NoStop}%
\bibitem [{\citenamefont {Giannozzi}\ \emph {et~al.}(2017)\citenamefont
  {Giannozzi} \emph {et~al.}}]{Giannozzi2017}%
  \BibitemOpen
  \bibfield  {author} {\bibinfo {author} {\bibfnamefont {P.}~\bibnamefont
  {Giannozzi}} \emph {et~al.},\ }\bibfield  {title} {\bibinfo {title} {Advanced
  capabilities for materials modelling with quantum espresso},\ }\href
  {https://doi.org/10.1088/1361-648X/aa8f79} {\bibfield  {journal} {\bibinfo
  {journal} {J. Phys.: Condens. Matter}\ }\textbf {\bibinfo {volume} {29}},\
  \bibinfo {pages} {465901} (\bibinfo {year} {2017})}\BibitemShut {NoStop}%
\bibitem [{\citenamefont {Prandini}\ \emph {et~al.}(2018)\citenamefont
  {Prandini}, \citenamefont {Marrazzo}, \citenamefont {Castelli}, \citenamefont
  {Mounet},\ and\ \citenamefont {Marzari}}]{Prandini2018}%
  \BibitemOpen
  \bibfield  {author} {\bibinfo {author} {\bibfnamefont {G.}~\bibnamefont
  {Prandini}}, \bibinfo {author} {\bibfnamefont {A.}~\bibnamefont {Marrazzo}},
  \bibinfo {author} {\bibfnamefont {I.~E.}\ \bibnamefont {Castelli}}, \bibinfo
  {author} {\bibfnamefont {N.}~\bibnamefont {Mounet}},\ and\ \bibinfo {author}
  {\bibfnamefont {N.}~\bibnamefont {Marzari}},\ }\bibfield  {title} {\bibinfo
  {title} {Precision and efficiency in solid-state pseudopotential
  calculations},\ }\href {https://doi.org/10.1038/s41524-018-0127-2} {\bibfield
   {journal} {\bibinfo  {journal} {npj Computational Materials}\ }\textbf
  {\bibinfo {volume} {4}},\ \bibinfo {pages} {72} (\bibinfo {year}
  {2018})}\BibitemShut {NoStop}%
\bibitem [{\citenamefont {Monkhorst}\ and\ \citenamefont
  {Pack}(1976)}]{Monkhorst1976}%
  \BibitemOpen
  \bibfield  {author} {\bibinfo {author} {\bibfnamefont {H.~J.}\ \bibnamefont
  {Monkhorst}}\ and\ \bibinfo {author} {\bibfnamefont {J.~D.}\ \bibnamefont
  {Pack}},\ }\bibfield  {title} {\bibinfo {title} {Special points for
  brillouin-zone integrations},\ }\href
  {https://doi.org/10.1103/PhysRevB.13.5188} {\bibfield  {journal} {\bibinfo
  {journal} {Phys. Rev. B}\ }\textbf {\bibinfo {volume} {13}},\ \bibinfo
  {pages} {5188} (\bibinfo {year} {1976})}\BibitemShut {NoStop}%
\bibitem [{\citenamefont {Marzari}\ \emph {et~al.}(2012)\citenamefont
  {Marzari}, \citenamefont {Mostofi}, \citenamefont {Yates}, \citenamefont
  {Souza},\ and\ \citenamefont {Vanderbilt}}]{Marzari2012}%
  \BibitemOpen
  \bibfield  {author} {\bibinfo {author} {\bibfnamefont {N.}~\bibnamefont
  {Marzari}}, \bibinfo {author} {\bibfnamefont {A.~A.}\ \bibnamefont
  {Mostofi}}, \bibinfo {author} {\bibfnamefont {J.~R.}\ \bibnamefont {Yates}},
  \bibinfo {author} {\bibfnamefont {I.}~\bibnamefont {Souza}},\ and\ \bibinfo
  {author} {\bibfnamefont {D.}~\bibnamefont {Vanderbilt}},\ }\bibfield  {title}
  {\bibinfo {title} {Maximally localized wannier functions: Theory and
  applications},\ }\href {https://doi.org/10.1103/RevModPhys.84.1419}
  {\bibfield  {journal} {\bibinfo  {journal} {Rev. Mod. Phys.}\ }\textbf
  {\bibinfo {volume} {84}},\ \bibinfo {pages} {1419} (\bibinfo {year}
  {2012})}\BibitemShut {NoStop}%
\bibitem [{\citenamefont {Mostofi}\ \emph {et~al.}(2014)\citenamefont {Mostofi}
  \emph {et~al.}}]{Mostofi2014}%
  \BibitemOpen
  \bibfield  {author} {\bibinfo {author} {\bibfnamefont {A.~A.}\ \bibnamefont
  {Mostofi}} \emph {et~al.},\ }\bibfield  {title} {\bibinfo {title} {An updated
  version of wannier90: A tool for obtaining maximally-localised wannier
  functions},\ }\href {https://doi.org/10.1016/j.cpc.2014.05.003} {\bibfield
  {journal} {\bibinfo  {journal} {Comput. Phys. Commun.}\ }\textbf {\bibinfo
  {volume} {185}},\ \bibinfo {pages} {2309–2310} (\bibinfo {year}
  {2014})}\BibitemShut {NoStop}%
\bibitem [{\citenamefont {Rourke}\ and\ \citenamefont
  {Julian}(2012)}]{Rourke2012}%
  \BibitemOpen
  \bibfield  {author} {\bibinfo {author} {\bibfnamefont {P.~M.~C.}\
  \bibnamefont {Rourke}}\ and\ \bibinfo {author} {\bibfnamefont {S.~R.}\
  \bibnamefont {Julian}},\ }\bibfield  {title} {\bibinfo {title} {Numerical
  extraction of de haas–van alphen frequencies from calculated band
  energies},\ }\href {https://doi.org/10.1016/j.cpc.2011.10.015} {\bibfield
  {journal} {\bibinfo  {journal} {Computer Physics Communications}\ }\textbf
  {\bibinfo {volume} {183}},\ \bibinfo {pages} {324–332} (\bibinfo {year}
  {2012})}\BibitemShut {NoStop}%
\bibitem [{\citenamefont {Yang}\ \emph {et~al.}(2018)\citenamefont {Yang},
  \citenamefont {Moessner},\ and\ \citenamefont {Lim}}]{Yang2018}%
  \BibitemOpen
  \bibfield  {author} {\bibinfo {author} {\bibfnamefont {H.}~\bibnamefont
  {Yang}}, \bibinfo {author} {\bibfnamefont {R.}~\bibnamefont {Moessner}},\
  and\ \bibinfo {author} {\bibfnamefont {L.-K.}\ \bibnamefont {Lim}},\
  }\bibfield  {title} {\bibinfo {title} {Quantum oscillations in nodal line
  systems},\ }\href {https://doi.org/10.1103/PhysRevB.97.165118} {\bibfield
  {journal} {\bibinfo  {journal} {Phys. Rev. B}\ }\textbf {\bibinfo {volume}
  {97}},\ \bibinfo {pages} {165118} (\bibinfo {year} {2018})}\BibitemShut
  {NoStop}%
\bibitem [{\citenamefont {Oroszlány}\ \emph {et~al.}(2018)\citenamefont
  {Oroszlány}, \citenamefont {Dóra}, \citenamefont {Cserti},\ and\
  \citenamefont {Cortijo}}]{Oroszlany2018}%
  \BibitemOpen
  \bibfield  {author} {\bibinfo {author} {\bibfnamefont {L.}~\bibnamefont
  {Oroszlány}}, \bibinfo {author} {\bibfnamefont {B.}~\bibnamefont {Dóra}},
  \bibinfo {author} {\bibfnamefont {J.}~\bibnamefont {Cserti}},\ and\ \bibinfo
  {author} {\bibfnamefont {A.}~\bibnamefont {Cortijo}},\ }\bibfield  {title}
  {\bibinfo {title} {Topological and trivial magnetic oscillations in nodal
  loop semimetals},\ }\href {https://doi.org/10.1103/PhysRevB.97.205107}
  {\bibfield  {journal} {\bibinfo  {journal} {Phys. Rev. B}\ }\textbf {\bibinfo
  {volume} {97}},\ \bibinfo {pages} {205107} (\bibinfo {year}
  {2018})}\BibitemShut {NoStop}%
\end{thebibliography}
%

\end{document}